\documentclass[a4paper,UKenglish,cleveref, autoref, thm-restate]{lipics-v2021}

\hideLIPIcs  

\title{Incremental computation of the set of period sets} %

\author{Eric Rivals}{LIRMM, Université Montpellier, CNRS, Montpellier, France \and \url{https://www.lirmm.fr}}{rivals@lirmm.fr}{https://orcid.org/0000-0003-3791-3973}{}
\authorrunning{E. Rivals} 

\Copyright{Eric Rivals} 


\relatedversion{A version of this paper is accepted for publication at the conference \href{http://www.sofsem.sk/}{SOFSEM 2025} and is also available on HAL repository at \url{https://hal-lirmm.ccsd.cnrs.fr/lirmm-04531880}.}

\supplement{\textbf{Resource}: we provide files containing the period sets of $\Gn$ for $n = 1, \ldots, 100$ on Zenodo at \url{https://doi.org/10.5281/zenodo.13826259} ---\cite{rivals-gamma-data}.
}

\funding{E. Rivals is supported by the European Union’s Horizon 2020 research and innovation programme under the Marie Skłodowska-Curie grant agreement No 956229.}

\nolinenumbers 

\EventEditors{???}
\EventNoEds{3}
\EventLongTitle{???}
\EventShortTitle{2024}
\EventAcronym{???}
\EventYear{2024}
\EventDate{June xx--xx, 2024}
\EventLocation{xx, xx}
\EventLogo{}
\SeriesVolume{}
\ArticleNo{}

\usepackage{amssymb,amsmath}
\usepackage[table]{xcolor}
\usepackage{pslatex}
\usepackage{subcaption}
\usepackage{xcolor} 
\usepackage{colortbl}
\usepackage[ruled,vlined]{algorithm2e}
\usepackage{float}

\newcommand{\card}[1]{\sharp (#1)}

\newcommand{\length}[1]{\vert #1 \vert}

\newcommand{\zon}{{{\{0,1\}}^{n}}}

\newcommand{\Ss}{{{\Sigma}^{*}}}
\newcommand{\Sn}{{{\Sigma}^{n}}}

\newcommand{\Kn}{{\kappa_n}}
\newcommand{\Knp}{{\kappa_{n,p}}}

\newcommand{\Gn}{{\ensuremath{\Gamma_n}}}

\newcommand{\Gnm}{{\ensuremath{\Gamma_{n-1}}}}

\def\dd{\mathinner{.\,.}}

\DeclareMathOperator{\rfw}{rfw}

\renewcommand{\topfraction}{0.84}
\renewcommand{\bottomfraction}{0.6}
\renewcommand{\textfraction}{0.1}
\renewcommand{\floatpagefraction}{0.70095}
\usepackage{hyperref}
\hypersetup{
  final,pdftex,colorlinks=true,
  citecolor=blue,
  linkcolor=magenta!60!black,
  urlcolor=green!50!black,
  pdftitle={Incremental computation of the set of period sets},
  pdfauthor={Eric Rivals},
  pdfkeywords={string, overlap, autocorrelation, combinatorics, dynamics},
  pdfsubject={computer science, combinatorics}}

\begin{document}

\maketitle              
\begin{abstract}
  Overlaps between words are crucial in many areas of computer science, such as code design, stringology, and bioinformatics.  A self overlapping word is characterized by its periods and borders. A period of a word $u$ is the starting position of a suffix of $u$ that is also a prefix $u$, and such a suffix is called a border.  Each word of length, say $n>0$, has a set of periods, but not all combinations of integers are sets of periods. Computing the period set of a word $u$ takes linear time in the length of $u$. We address the question of computing, the set, denoted $\Gamma_n$, of all period sets of words of length $n$.  Although period sets have been characterized, there is no formula to compute the cardinality of $\Gamma_n$ (which is exponential in $n$), and the known dynamic programming algorithm to enumerate $\Gamma_n$ suffers from its space complexity. We present an incremental approach to compute $\Gamma_n$ from $\Gamma_{n-1}$, which reduces the space complexity, and then a constructive certification algorithm useful for verification purposes. The incremental approach defines a parental relation between sets in $\Gamma_{n-1}$ and $\Gamma_n$, enabling one to investigate the dynamics of period sets, and their intriguing statistical properties. Moreover, the period set of a word $u$ is key for computing the absence probability of $u$ in random texts. Thus, knowing $\Gamma_n$ is useful to assess the significance of word statistics, such as the number of missing words in a random text.

\end{abstract}


\section{Introduction}\label{sec:intro}
Considering finite words, we say that a word $u$ overlaps a word $v$ if a suffix of $u$ equals a prefix of $v$ of the same length. A pair of words $u,v$ can have several overlaps of different lengths. For instance, over the alphabet $\{a, b\}$, let $u := ababba$ and $v:= abbabb$, then $u$ overlaps $v$ with the suffix-prefix $abba$, and with $a$. It appears that the longest overlap contains all other overlaps: to find all overlaps from $u$ to $v$, it suffices to study the overlaps of $abba$ with itself. 

For a word $u$, a suffix that equals a prefix of $u$ is called a \emph{border}, and the length of $u$ minus the length of a border, is called a \emph{period}.  Computing all self-overlaps of a word $u$ is computing all its borders or all its periods, which can be done in linear time (see the algorithm for computing the border array in~\cite[Chap. 1]{Smyth-book-03}).  This well-studied problem was also solved for preprocessing the pattern in the seminal Knuth-Morris-Pratt pattern matching algorithm~\cite{Knuth-Morris-Pratt-pm}): the borders serve to optimally shift the window along the text when seeking the pattern. 
For instance the word $ababaaba$ has length $n=8$ and set of periods: $\{0, 5, 7\}$ (zero being the trivial period - the whole word matches itself).

One can easily see that distinct words of the same length can share the same set of periods, even if one forbids a permutation of the alphabet. For a word $u$, let us denote by $P(u)$ its period set (which we abbreviate by PS). In this work, we investigate algorithms to enumerate all possible period sets for any words of a given length $n$. This set is denoted $\Gn$ for $n>0$ and is non trivial if the alphabet contains at least two symbols. Brute force enumeration can consider all possible words of length $n$ and compute their period set, but this approach becomes computationally unaffordable for $n>30$. Currently, only a dynamic programming algorithm exists to enumerate $\Gn$, but it suffers from high space complexity~\cite{rivals_jcta_2003}.

The notion of overlap in strings is crucial in many areas and applications, among others: combinatorics, bioinformatics, code design, or string algorithms. Interest in $\Gn$ sparkled mostly in the 80's, when researchers started to evaluate the average behavior of pattern matching algorithms, or that of filtering strategies for sequence alignment, text comparisons or clustering.
A powerful filtering when comparing two texts, is to list their $k$-long substrings (a.k.a. $k$-mers), for appropriate values of $k$, and then compute e.g. a Jaccard distance between their set of $k$-mers, to see whether the two texts are similar enough to warrant a costly alignment procedure~\cite{Ukkonen-qgram-TCS}.

In a different area, testing Pseudo-Random Number Generators can also be translated into a question on vocabulary statistics. Indeed, in random real numbers written as sequence of digits, all substrings of a given length, say $k$, should ideally have an almost equal number of occurrences (i.e., should not significantly deviate from the theoretical expectation).  Empirical tests, named \emph{Monkey Tests}, were developed for such generators~\cite{MAR-ZAM-1993,PER:WHI:1995,Leopardi:2009}.  It turns out that the absence probability of a word/string in a random text is essentially controlled by the period set of the word~\cite{GuOd81}. Hence, the need for enumerating $\Gn$ appears in diverse domains of the literature~\cite{C_cpm_common_words_2000,rahmann_cpc_2003}. 

In network communication, the so-called prefix-free, bifix-free, and cross-bifix-free codes are used for synchronization purposes. Their design require to select a set of words that are mutually non overlapping -- a long studied topic. Nielsen published in 1973 a construction algorithm~\cite{Nielsen_bifix_1973} together with a note on the expected duration of a search for a fixed pattern in random data~\cite{Nielsen_expect_search_1973}, thereby linking explicitly pattern matching and code design.  Improved algorithms for such code design were recently published~\cite{Bilotta,Bajic}.

Many aspects of periodicity of words were and are still extensively studied in combinatorics, e.g., \cite{Ehrenfeucht_1979,gabric_etal_inequality_periods_2021}. For instance, the periods of random words were investigated in~\cite{holub_shallit_random_words} and the Fine and Wilf (FW) Theorem was generalized to three or more words~\cite{Castelli_1999}. 
Algorithms for constructing extremal FW-words relative to a subset of periods were gradually improved in~\cite{Tijdeman_2003,Tijdeman_2009}, a question that is related to the ones we investigate in Section~\ref{sec:fate}.
Last, the sequence formed by the cardinality of $\Gn$, which is denoted $\Kn$, has an entry in the OEIS, and some known lower and upper bounds~\cite{GuOd81,rivals_jcta_2003}, which helped closing a conjecture related to  $\Kn$ in 2023~\cite{C_icalp_2023}.  

\textbf{Plan}. In Section~\ref{sec:prelim}, we introduce a notation, definitions, and review some known results.  In Section~\ref{sec:inc}, we propose an incremental approach to compute $\Gn$ from $\Gnm$, which uses $O(n)$ memory. In Section~\ref{sec:constructive}, we give an algorithm to build a word that is a witness for a period set.  In Section~\ref{sec:fate}, we propose to explore the dynamics of period sets when $n$ increases, before exploring the distribution of period sets according to their basic period and concluding in Section~\ref{sec:explore}.

\section{Related works, notation and preliminary results.}
\label{sec:prelim}

\subsection{Notation}\label{sec:notation}

Here we introduce a notation and basic definitions.

For two integers $p, q \in \mathbb{N}_{> 0}$, we denote the fact that $p$ divides $q$ by $p \mid q$ and the opposite by $p \nmid q$.
We consider that strings and arrays are indexed from $0$. We use $=$ to denote equality, and $:=$ to denote a definition.
The cardinality of a discrete set $X$ is denoted by $\card{X}$.

An \emph{alphabet} $\Sigma$ is a finite set of \emph{letters}. A finite sequence of elements of $\Sigma$ is called a \emph{word} or a \emph{string}. The set of all words over $\Sigma$ is denoted by $\Sigma^\star$, and $\varepsilon$ denotes the empty word.  For a word $x$, $\length{x}$ denotes the \emph{length} of $x$.  Let $n$ be an integer.  The set of all words of length $n$ is denoted by $\Sn$. Given two words $x$ and $y$, we denote by $x.y$ the \emph{concatenation} of $x$ and $y$. For a word $w$, we define the powers of $w$ inductively:  $w^0 := \epsilon$ and for any $n>0$, $w^n := w.w^{n-1}$.  A word $u$ is \emph{primitive} if there exists no word $v$ such that $u = v^k$ with $k \geq 2$.


Let $u := u[0 \dd n-1] \in \Sigma^n$. For any $0\leq i \leq j \leq n - 1$, we denote by $u[i-1]$ the $i^{th}$ symbol in $u$, and by $u[i \dd j]$, the substring starting at position $i$ and ending at position $j$. In particular, $u[0 \dd j]$ denotes a prefix and $u[i \dd n-1]$ a suffix of $u$.

Let $x,y \in \Ss$ and let $j$ be an integer such that $0 \leq j \leq \length{x}$ and $j \leq \length{y}$. If $x[n-j \dd  \length{x}-1] = y[0 \dd j-1]$, then the \emph{merge} of $x$ and $y$ with offset $j$, which is denoted by $x \oplus_j y$, is defined as the concatenation of $x[0 \dd n-j-1]$ with $y$. I.e., $x \oplus_j y := x[0 \dd n-j-1].y$.

\subsection{Periods, period set, and the set of period sets}\label{sec:periods}

Let us first define the concepts of period, period set, basic period, and set of period sets, and then recall some useful results. 

\begin{definition}[Period/border]\label{def:period1}
  The string $u = u[0 \dd n-1]$ has period $p \in \{0, 1, \ldots, n-1\}$ if and only if $u[0\dd n-p-1] = u[p\dd n-1]$, i.e., for all $0 \leq i \leq n-p-1$, we have $u[i] = u[i+p]$. Moreover, we consider that $p = 0$ is a period of any string of length $n$.
  The substring $u[p \dd n-1]$ is called a \emph{border}.
\end{definition}

Zero is also called the trivial period. The \emph{period set} of a string $u$ is the set of all its periods and is denoted by $P(u)$. Formally, for any length $n>0$, any word $u$ in $\Sn$,  $P(u) := \{ p \in \mathbb{N} : p \text{ is a period in } u\}$. The \emph{weight} of a period set is its cardinality. The smallest non-zero period of $u$ is called its \emph{basic period}. When $P(u) = \{ 0 \}$, we consider that its basic period is the string length $n$.

Note that two definitions of period set exist in the literature: the one above, also used in \cite{GuOd81} (even if the set was encoded in a binary string called autocorrelation), where $P(u)$ is a subset of $\{0, 1, \ldots, n-1\}$, and the one from \cite{Lothaire-ACW} in which it equals $P(u) \cup \{n\}$ (where the length $n$ is included in $P(u)$). We will use the first one.

Let $\Gn := \{ Q \subset  \{0, 1, \ldots, n-1\} \,|\, \exists u \in \Sn: Q = P(u) \}$ be the set of all period sets of strings of length  $n$. We denote its cardinality by $\Kn$. The period sets in $\Gn$ can be partitioned according to their basic period; thus, for $0\leq p < n$, we denote by $\Gamma_{n,p}$ the subset of period sets whose basic period is $p$, and by $\Knp$ the cardinality of this subset.
A surprising result from Guibas and Odlyzko's characterization of period sets~\cite{GuOd81} is the \emph{alphabet independence} of $\Gn$: Any alphabet of size at least two gives rise to the same set of period sets, i.e., to $\Gn$. In the sequel, we assume $\card{\Sigma} > 1$.   Circa 20 years later, Halava et al. gave a simpler proof of the alphabet independence result by solving the following question \cite{Halava:etal:2000}. For any period set $Q$ of $\Gn$, let $v$ be a word over an alphabet $\Sigma$ with $\card{\Sigma}>1$ such that $P(v)=Q$; then compute a binary word $u$ such that $P(u)=Q$. Indeed, they gave a linear time algorithm to compute such a word $u$ from $v$.

For the most important properties on periods, we refer the reader to~\cite{GuOd81,Lothaire-COW-97,Lothaire-ACW} and Appendix~\ref{app:periods}. Below we recall the two characterizations of period sets from \cite{GuOd81} and from \cite{Lothaire-ACW}, and state the version for strings of the famous Fine and Wilf (FW) Theorem~\cite{FW65}. We refer the reader to \cite{rivals_jcta_2003,C_icalp_2023} for more properties on period sets and on $\Gn$.

\subsection{Characterizations of period sets}
\label{sec:charac}

For a given word length $n>0$, the question of characterization is: among all possibles subsets of $\{0, 1, \ldots, n-1\}$, recognize those that are period sets for at least one word of length $n$.

In 1981, Guibas and Odlyzko (GO for short) proved a double characterization of period sets: one is based on two rules, the Forward Propagation Rule (FPR) and the Backward Propagation Rule (BPR), the second is the recursive predicate \(\Xi\) (which is given in extenso in Appendix~\ref{sec:xi}).

First, let us formulate the FPR and BPR in terms of sets (rather than in term of binary vector as in \cite{GuOd81}). 
Let \(P\) a subset of \(\{0, 1, \ldots, n-1 \}\). 
\begin{definition}[FPR]
\(P\) satisfies the FPR iff for all pairs \(p,q\) in \(P\) satisfying \(0\leq p < q < n\), it follows that \(p+i(q-p) \in P\) for all \(i=2,\ldots,\lfloor (n-p)/(q-p) \rfloor\).
\end{definition}

\begin{definition}[BPR]
\(P\) satisfies the BPR iff for all pairs \(p,q\) in \(P\) satisfying \(0 \leq p < q < 2p\), and \((2p-q) \not\in P\), it follows that
\(p-i(q-p) \not\in P\) for all \(i=2,\dots,\min(\lfloor p/(q-p) \rfloor, \lfloor (n-p)/(q-p) \rfloor)\).
\end{definition}

We can now state GO's characterization theorem
\begin{theorem}\label{th:go:charac}
  Let \(P\) a subset of \(\{0, 1, \ldots, n-1 \}\). The four following statements are equivalent:
  (1) $P$ is the period set of a binary word of length $n$.\\
  (2) $P$ is the period set of a word of length $n$.\\
  (3) Zero belongs to $P$ and $P$ satisfies the forward and backward propagation rules.\\
  (4) $P$ satisfies predicate $\Xi$.
\end{theorem}
The equivalence of (1) and (2) yields the abovementioned alphabet independence of $\Gn$. 
The authors also noticed that the FPR and BPR are local properties~\cite[Lemma 3.1]{GuOd81}, which we use later on.
Finally, predicate $\Xi$ is a recursive procedure to check whether $P$ belongs to $\Gn$ in $O(n)$ time: it considers two cases depending on whether the basic period is $\leq \lfloor n/2\rfloor$ (case a) or not (case b)  (cf. Appendix~\ref{sec:xi}).

Besides this characterization, \cite{GuOd81} studied the set of distinct strings over a given alphabet that share the same period set (a.k.a. its \emph{population}). They also reported values of $\Kn$, the cardinality of $\Gn$, for $n < 55$, exhibited lower and upper bounds for $\log(\Kn)/\log(n)^2$, and conjectured its convergence when $n \rightarrow \infty$. Much later, improved lower and upper bounds were provided and served to close the conjecture \cite{rivals_jcta_2003,C_icalp_2023}.

An alternative characterization of PS appears in 2002 as Theorem 8.1.11 in \cite[Chap. 8]{Lothaire-ACW}; it comprises four equivalent statements, of which the first three "are proved in~\cite{GuOd81}" (see notes in \cite[p. 310]{Lothaire-ACW}; the first two are identical to (1) and (2) in Theorem~\ref{th:go:charac}). We thus recall only statement (iv) of this characterization in the next theorem.

\begin{theorem}[\protect\cite{Lothaire-ACW}]\label{th:acw:charac}
Let $P := \{0 = p_0 < p_1 < \ldots < p_s = n\}$ be a set of integers and let $d_h := p_h - p_{h-1}$ for $1 \leq h \leq s$.
Then $P$ is a period set (i.e., $P \in \Gamma_n$) if and only if for each $h$ such that $d_h + p_h \leq n$, one has: 
(a) $p_h + d_h \in P$ and (b) if $d_h = k d_{h+1}$ for some integer $k$, then $k=1$.
\end{theorem}
Note that in this formulation $n$ belongs to $P$ (see our remark about two definitions of period sets), but as when $h = s$ one has  $d_s + p_s > n$, one sees that conditions (a) and (b) are in fact only required for $1 \leq h < s$.

Given a period set $P(w)$ of some word $w$, the proof shows that there exists a binary word having the period set $P(w)$. It is a proof of existence, it is not constructive, but the authors also give an example on how to build a binary string for $Q := \{0, 11, 14, 17, 18 \}$ knowing that $Q$ is the period set of the word $w = abcabca defg abcabca$. Knowing that $Q \in \Gn$, an algorithm for solving this problem was given by~\cite{Halava:etal:2000}.

In this work, we address a different question, termed \emph{constructive certification}: Given \(Q\) a subset of \(\{0, 1, \ldots, n-1 \}\), build a (binary) string $u$ such that $P(u) = Q$ iff $Q \in \Gn$, or return the empty string otherwise.

\subsection{Additional properties of periods}
\label{sec:rw:periods}

In \cite{GuOd81}, the authors give a version for strings of the famous Fine and Wilf Theorem \cite{FW65}, a.k.a. the periodicity lemma. A nice proof was provided by Halava and colleagues \cite{Halava:etal:2000}.
\begin{theorem}[Fine and Wilf]\label{thm:gcd}
Let $p, q$ be periods of $u \in \Sigma^n$. If $n \geq p + q - \gcd(p,q)$, then $\gcd(p,q)$ is a period of $u$.
\end{theorem}
\noindent
We can reformulate Theorem~\ref{thm:gcd} as a condition that must be satisfied by a period set \(P\).
\begin{theorem}\label{thm:gcd:cond}
Let $P \in \Gn$. Let \(p, q\) be periods of \(P\) such that \(\gcd(p,q) \not\in P\), and define \(FW(p,q) := p + q - \gcd(p,q)\).
Then \(FW(p,q)\) must be strictly larger than \(n\).
\end{theorem}
We call \(FW(p,q)\) the \emph{Fine and Wilf (FW) limit} of \((p,q)\). We say $(p,q)$ violates the \emph{FW condition} at length $n$, if \(FW(p,q) \leq n\).

\subsection{Dynamic programming algorithm to enumerate $\Gn$}
\label{sec:dyn:prog}

In 2001, further investigation of $\Gn$ led to a dynamic programming programming algorithm to enumerate all period sets in $\Gn$: it converts the recursive approach of predicate $\Xi$ into a dynamic program that stores all $\Gamma_i$ for $0 < i \leq \lfloor 2n/3 \rfloor$~\cite{C_icalp_period_2001,rivals_jcta_2003}. With some practical improvements, the range was reduced to $0 < i \leq \lfloor n/2 \rfloor$. However, as $\Kn$ is exponential in $n$, this induces a large memory usage, which remains a serious drawback. Hence, the quest for memory efficient algorithms. The authors also demonstrated that $\Gn$ equipped with set inclusion is a lattice, but this did not help to improve $\Gn$ enumeration.


\section{Incremental enumeration of $\Gn$}
\label{sec:inc}

\textbf{Rationale of the incremental approach}
\label{sec:org19f7988}\label{sec:inc:rationale}

In their seminal work~\cite{GuOd81}, GO manipulate period sets, not as sets, but as a binary strings of length $n$, where position $i$ is set to \texttt{1} if $i$ is a period. For example, the binary encoding of period set $\{0, 3, 5\}$ for length $n=6$ is \texttt{100101}. They call these binary strings correlations, and even autocorrelations to emphasize that it encodes all self-overlaps a string ~\cite[p. 21]{GuOd81}. The rule based characterization (statement (3) of Theorem~\ref{th:go:charac}) implies a \emph{special substring} property \cite{GuOd81}. Indeed, noting the locality of the forward and backward propagation rules, the authors state in Lemma 3.1 in \cite{GuOd81}: If a binary string \(v\) satisfies the forward and backward propagation rules, then so does any prefix or suffix of \(v\).

As the rule based characterization of period sets -- statement (3) of Theorem~\ref{th:go:charac} -- also requires that an autocorrelation has its first bit equal to one, or equivalently that zero belongs to a period set, one gets the following theorem (\cite[Thm 1.3]{rivals_jcta_2003}):
\begin{theorem}\label{thm:ac:sub}
Let \(v\) be an autocorrelation of length \(n\). Any substring \(v_i\dots v_j\) of \(v\) with \(0 \leq i \leq j < n\) such that \(v_i = 1\) is an autocorrelation of length \(j-i+1\).
\end{theorem}
Applying Theorem~\ref{thm:ac:sub} to a prefix of $v$, one gets for any \(n>0\): The prefix of length \((n-1)\) from an autocorrelation of length \(n\) is an autocorrelation of length \((n-1)\).
In terms of period sets, this statement can be reformulated as:
\begin{corollary}
If \(P\) is a period set of \(\Gn\), then \(P \setminus \{ n-1 \}\) belongs to \(\Gnm\).
\end{corollary}
First, this means that, knowing \(\Gn\), it is easy to compute \(\Gnm\). It suffices to consider each element of \(\Gn\) in turn (or in parallel) and to possibly remove the period \((n-1)\) from it  (i.e., if \((n-1)\) belongs to it) to obtain an element of \(\Gnm\).  With this procedure one can obtain the same element of \(\Gnm\) twice, and one must keep track of this to avoid redundancy.

\noindent
Conversely, we get the Lemma that underlies the incremental approach for computing \(\Gn\):
\begin{lemma}\label{lem:gnm:gn}\label{cor:gnm:gn}
Let \(Q\) be a period set of \(\Gn\). Then $Q$ can only be of two alternative forms: either \(P\) or \(P \cup \{ n-1 \}\), for some $P$ in \(\Gnm\).
\end{lemma}

\noindent
\textbf{Incremental algorithm framework}
\label{sec:org4693e8e}\label{sec:inc:frame}

Lemma~\ref{cor:gnm:gn} suggests an approach for computing \(\Gn\) using \(\Gnm\). Consider each \(P\) from \(\Gnm\), and check whether the candidate sets \(P\) and \(P \cup \{ n-1 \}\) are period sets of \(\Gn\).  Algorithm \ref{algo:inc:gen} presents a generic incremental algorithm for \(\Gn\), where \texttt{certify} denotes the certification function used. In general, a certification function takes as input \(n\) and any subset $Q$ of \(\{0, 1, \ldots, n-1 \}\), returns True if and only if $Q$ belongs to \(\Gn\).

\begin{algorithm}[htbp]
  \SetKwInOut{Input}{Input}
  \caption{\label{algo:inc:gen} IncrementalGamma( length $n>1$; set $\Gamma_{n-1}$ )}
  \textbf{Output}: {$\Gamma_n$: the set of period sets for length $n$}\;
  \BlankLine
  \nl $G := \emptyset$\tcp*{   $G$: variable to store $\Gamma_n$}
  \nl \For{all $P \in \Gamma(n-1)$ }{
    \nl \lIf{\texttt{certify}( $P$, $n$ )}{ insert $P$ in $G$ }
    \nl  $Q := P \cup \{ n-1 \}$ \tcp*[h]{build extension $P$ with period $n-1$}\;
    \nl \lIf{\texttt{certify}( $Q$, $n$ )}{ insert $Q$ in $G$ }
  }
  \nl   \KwRet{$G$}\;
\end{algorithm}

The recursive predicate \(\Xi\) from \cite{GuOd81} (cf. Appendix~\ref{sec:xi}) is indeed a certification function: it does exactly what is required for any subset of \(\{0, 1, \ldots, n-1\}\), in $O(n)$ time \cite{GuOd81}. Thus, using predicate $\Xi$, Algorithm~\ref{algo:inc:gen} correctly computes \(\Gn\) from \(\Gnm\). We will discuss alternative certification functions below.

Besides its simplicity, the main advantage of Algorithm~\ref{algo:inc:gen}, compared to the dynamic programming enumeration algorithm of \cite{C_icalp_period_2001}, is its space complexity. Here, the computation considers each period set \(P\) from \(\Gnm\) in turn (and independently from the others), executes twice the certification function for \(P\) and \(Q\); this implies that the memory required, besides storage of \(P, Q\), is the one used by the certification function. With the predicate \(\Xi\), it is linear in \(n\), so \(O(n)\) space. The time complexity is proportional to \(\Kn\) (i.e., the cardinality of \(\Gn\)) times the running time of the certification function, which yields the following theorem. Of course, the set $\Gnm$ must be available on disk space before hand. 

\begin{theorem}\label{th:inc:gen}
  Provided that $\Gnm$ is available on external memory, then
  \begin{enumerate}
  \item Algorithm \ref{algo:inc:gen} using any certification function correctly computes \(\Gn\) from \(\Gnm\).
  \item Using the predicate \(\Xi\) as certification function, it runs in \(O(n\Kn)\) time and \(O(n)\) space.
  \end{enumerate}
\end{theorem}
Moreover, it is worth noticing that Algorithm~\ref{algo:inc:gen} is embarrassingly parallelizable.
Note that the output contains $\Kn$ period sets, whose cardinality is bounded by $n$ and sometimes equal $n$. For known bounds on $\Kn$ see~\cite{rivals_jcta_2003,C_icalp_2023}. So, the output size is bounded by $n\Kn$.

\medskip
\noindent
\textbf{Alternative certification functions}
\label{sec:org271be92}\label{sec:alt}

In our incremental setup, that is when computing $\Gn$ using  \(\Gnm\), we know that \(P\) belongs to \(\Gnm\) (line 2 in Algorithm~\ref{algo:inc:gen}). Hence, the candidate sets, denoted $P$ and $Q$, are not \textbf{any subset} of \(\{0, 1, \ldots, n-1 \}\), but already satisfy some constraints for length \((n-1)\).  Therefore, finding alternative certification functions is interesting.  Here, we discuss two alternative functions, and in Section~\ref{sec:constructive}, we exhibit a constructive certification algorithm, which not only certifies a candidate set, but also computes a witness, i.e., a word whose period set is the candidate set, only if the answer is positive.

To simplify Algorithm~\ref{algo:inc:gen} by improving how certifications are done for the candidate sets  $P$ and $Q := P \cup \{ n-1 \}$, we can take advantage of two facts. First, the only period that can be added is $n-1$. This limits the cases for which we need to check some conditions. Second, $P$ and $Q$ are not independent, and if $n-1$ is compulsory at length $n$, because it is generated by the FPR from smaller periods, then only candidate set $Q := P \cup \{ n-1 \}$ may belong to $\Gn$, but not $P$.

Besides Predicate $\Xi$, a second certification function, we call it \emph{rule based}, exploits statement (3) of Theorem~\ref{th:go:charac} and uses procedures to verify if the forward and backward propagation rules are satisfied. Algorithm~\ref{algo:inc} gives the code for computing \(\Gn\) using the rule based certification (see explanations in Appendix~\ref{sec:inc:rule}).

A third certification approach, which exploits the characterization of Theorem~\ref{th:acw:charac}, can perform the certification of both $P$ and $Q$ in $O(\card{P})$ time. Some details are also given in Appendix~\ref{sec:alt:acw}).

\section{Constructive certification of a period set}
\label{sec:constructive}


Let \(Q\) be subset of \(\{0, 1, \ldots, n-1\}\).  We say that a word \(u\) \emph{realizes} \(Q\) if \(P(u) = Q\).

The certification functions used in Section~\ref{sec:inc} yield a True/False answer, but no witness for a period set. As there is no resources providing $\Gn$ for many word lengths, checking the output of an enumeration algorithm, remains difficult. Hence, we need a constructive certification function, which given length $n>0$ and set $Q$, provides a word realizing $Q$ if only if $Q \in \Gn$, and the empty string otherwise.  Given the alphabet independence of \(\Gn\), we restrict the search to binary words. We present an algorithm called \emph{binary realization} (see Algorithm~\ref{algo:bin:real}) solving this question, and demonstrate its linear complexity. 

Using Algorithm~\ref{algo:bin:real} as a certification function in Algorithm~\ref{algo:inc:gen}, for each $R \in \Gn$, we get a word $u$ realizing $R$. Then, computing the period set of $u$ allows us to check that $P(u) = R$.
Let us define the notion of \emph{nested set}.
\begin{definition}\label{def:nested}
  Let \(n>0\), \(P\) be a subset of \(\{0, 1, \ldots, n-1\}\), and \(q\) be an element of \(P\). We denote by \(P_q\), the \emph{nested set of P starting at period} \(q\):
  \[ P_q := \{ (r-q) \text{ for each } r \in P \text{ such that } r \geq q \}.\]
\end{definition}
By construction \(P_q\) starts with \(0\); moreover, if we choose \(q=0\) then \(P_q = P\).

Another interesting certification function is: to attempt to build a word \(u\) that realizes \(R\); if the attempt succeeds, \(R\) is a valid period set. 
Below we present an algorithm for the \emph{binary realization} of a set (see Algorithm~\ref{algo:bin:real}).
Using it as a certification function in Algorithm~\ref{algo:inc:gen}, the latter will compute \(\Gn\) from \(\Gnm\) and also yield one realizing string for each period set.

\subsection{Binary realization of a subset of \(\{0, 1, \ldots, n-1\}\)}
\label{sec:orge1cfb4e}

Algorithm~\ref{algo:bin:real} computes a word \(u\) that realizes a set \(P\) for length \(n>0\), or returns the empty word \(\epsilon\) if \(P\) is not a period set of \(\Gn\).  For legibility, the preliminary checks on \(P\) are not written in Algorithm~\ref{algo:bin:real}: they include checking that \(P\) is a subset of \([0, \ldots, n-1]\), is ordered, and has zero as first period. The word \(u\) is written over the alphabet \(\{\mathtt{a}, \mathtt{b}\}\).
\begin{algorithm}
  \SetKwInOut{Input}{Input}
  \caption{\label{algo:bin:real} Binary Realization}
  \Input{$n>0$: integer; $P$: a subset of $[0,1, \ldots, n-1]$ including $0$, in a sorted array} 
  \textbf{Output}: {a \textbf{binary} string realizing $P$ at length $n$ xor the empty string otherwise}\;
  \BlankLine
  \nl $k := \card{P}$ \tcp*{   $k$: cardinality of  $P$}
  \nl \lIf{ $k = 1$ }{\KwRet{ $\mathtt{a.b^{(n-1)}}$ }\tcp*[f]{trivial case where $P=\{0\}$}}
  \tcp{processing the largest period and init. variables}
  \nl prevLg $:= n - P[k-1]$; prevIP $:= $ prevLg; prevSuffix $:= \mathtt{a.b^{(\text{prevLg} -1)}}$\;

  \nl \For{ $i$ going from $k-2$ to $0$ }{
      \nl lg $:= n - P[i]$\;
      \nl innerPeriod $:= P[i+1] - P[i]$\;
      \nl \lIf{ innerPeriod $<$ prevIP }{ \KwRet{ $\epsilon$ } }
     
      \nl \If(\tcp*[f]{condition for case 1}){lg $\leq 2 \times $ prevLg}{
        
          \nl \If{ ( innerPeriod $=$ prevIP ) OR (( prevIP $\nmid$ innerPeriod ) AND ( (innerPeriod $=$ prevLg) OR ( prevSuffix has period innerPeriod ) ) ) }{
            \tcp{suffix := a prefix of prevSuffix concat. with prevSuffix}
            \nl suffix $:=$ prevSuffix[$0..$innerPeriod$-1$] . prevSuffix\;
          }
          \nl \lElse{ \KwRet{ $\epsilon$ } \tcp*[f]{invalid case for length lg}}
       }
       \nl\Else(\tcp*[f]{condition for case 2}){
         \tcp{suffix := prevSuffix newsymbols prevSuffix}
         \nl m $:= $ lg $- 2 \times $ prevLg\;
         \nl newPrefix $:= $ prevSuffix $. \mathtt{a^{m}}$\;
         \nl \If{ \text{newPrefix} is not primitive }{
           \nl newPrefix $:= $ prevSuffix $. \mathtt{a^{(m-1)}b}$\;
         }
         \tcp{Invariant: newPrefix is primitive}
         \nl suffix $:=$ newPrefix . prevSuffix\;
       }
       \tcp{update variables}
       \nl prevLg $:= $ lg; prevIP $:= $ innerPeriod; prevSuffix $:= $ suffix\;
     }
     \nl \KwRet{ suffix }\;
\end{algorithm}

The algorithm considers elements of \(P\) backwards, starting with largest integer first, since \(P\) is ordered. At each execution of the for loop, it considers the current integer \(P[i]\) as a period and builds a suffix of \(u\) of length \(n-P[i]\) (variable \emph{lg}). In fact, it considers a potentially larger and larger nested sets, and computes a suffix of \(u\) for this length. At the end of the for loop, the variable \emph{suffix} contains a string of length \emph{lg} realizing the nested set. Note that algorithm uses three variables (whose names start with \emph{prev}) to store the length, the inner period, and the suffix obtained with the previous period.

The base case is processed before the loop and consider the nested set for \(P[k-1] = max(P)\) for the length \(n-max(P)\) without any period. Hence, the suffix \(\mathtt{a}.\mathtt{b}^{(\text{prevLg -1})}\) is a realization for nested set $\{0\}$ for length \(n-max(P)\).

In the for loop the key variable is the \emph{innerPeriod}, which equals the offset \(P[i+1]-P[i]\), which is the basic period of the current nested set. If $\text{innerPeriod} < \text{prevIP}$ then the FPR is violated and the algorithm returns $\epsilon$ (line 7). Two cases are considered depending on whether the current length is smaller twice the previous length (case 1) or not (case 2). Because of the notion of period, the suffix must start and end with a copy of \emph{prevSuffix}. The construction of the suffix depends on the case. In case 1, the two copies of  \emph{prevSuffix} are concatenated or overlap themselves, and some additional conditions are required (line 9). These conditions are dictated by the characterization of period set from~\cite{GuOd81} (see the predicate \(\Xi\) in Appendix~\ref{sec:xi}).  Whenever one is not satisfied, Algorithm~\ref{algo:bin:real} returns the empty word as expected.  In case 2, the two copies of \emph{prevSuffix} must be separated by \emph{m} additional symbols (to be determined).  One builds a \emph{newPrefix} that starts with  \emph{prevSuffix} followed by \(\mathtt{a}^{m}\), and one checks whether \emph{newPrefix} is primitive. This \emph{newPrefix} is the part that ensures the suffix will have \emph{innerPeriod} as a period. The primitivity is required, since \emph{newPrefix} may have a proper period, but this period shall not divide \emph{innerPeriod}. If primitivity is not satisfied, then changing the last symbol $\mathtt{a}$ of \emph{newPrefix} by \(\mathtt{b}\) will make it primitive. This is enforced by Lemma 3 from~\cite{Halava:etal:2000}, which states that for any binary word \(w\), \(w\mathtt{a}\) or \(w\mathtt{b}\) is primitive. So, we know that at least one of the two forms of \emph{newPrefix} is primitive as necessary. It can be that both are primitive and suitable. Finally, we build the current \emph{suffix} by concatenating \emph{newPrefix} with \emph{prevSuffix}.

\textbf{Complexity}  First, in case 1, checking the condition "\emph{prevSuffix has period innerPeriod}" can be done in linear time in $\length{\mathit{prevSuffix}}$ (which is $\leq n$). Overall, this can be executed $\card{P}$ times.  Second, the primitivity test performed in case 2 takes a time proportional to the length of the string \emph{newPrefix}. However, the sum of these lengths, for all iterations of the loop, is bounded by \(n\).  Other instructions of the for loop take constant time. Overall, the time complexity of Algorithm~\ref{algo:bin:real} by \(O(\card{P} \times n)\). However, when Algorithm~\ref{algo:bin:real} is plugged in Algorithm~\ref{algo:inc:gen} it processes special instances: either $P$ or $Q := P \cup \{ n-1 \}$, with $P \in \Gnm$. Then, the time taken by all verifications of condition "\emph{prevSuffix has period innerPeriod}" for all cases 1 is bounded by $n$, due to the properties of periods that generate more than two repetitions in a string (see Lemma~\ref{lem:multiply} and Lemma 2 from~\cite{Halava:etal:2000}). For the instances processed in Algorithm~\ref{algo:inc:gen}, the time complexity of Algorithm~\ref{algo:bin:real} is \(O(\card{P} + n)\) or $O(n)$.

\textbf{Remark} we can modify Algorithm~\ref{algo:bin:real} to build, instead of a binary word, a realizing word that maximizes the number of distinct symbols used in it.
Indeed, new symbols are used only in the base case and in case 2. Each time, it is possible to choose symbols that have not been used earlier in the algorithm, and thus to maximize the overall number. Note that this would remove the need of the primitivity test in case 2.

\subsection{Examples of binary realization}
\label{sec:org555a1b0}\label{sec:bin:real}
We consider the case of the period set \(P := \{0,3,6,8 \}\) from $\Gamma_{9}$, which does not belong to $\Gamma_{10}$, and show the traces of execution for both lengths \(n:=9\) and \(n:=10\).
The table below shows the trace for \(n:=9\). The operator \({\oplus_j}\) merges the two strings with an offset of length \(j\) if the corresponding prefix and suffix are equal, for any appropriate integer \(j\). So when \(n=9\) and $i=0$, the merge \(v := w{\oplus_3}w\) with \(w = abaaba\) is feasible since \(w\) has period \(3\). When \(n=10\), the merge \(w := y{\oplus_3}y\) with \(y = abab\) is not possible since \(a \neq b\).

\begin{center}
  \begin{tabular}{>{\columncolor[gray]{.8}}clcc>{\columncolor[gray]{.8}}ll}
    \rowcolor[gray]{.8}
    period & length & inner period & case & suffix & valid\\
    \hline
    8 & 9-8 = 1 & 9-8 = 1 & 2 & \(z := a\) & true\\
    6 & 9-6 = 3 & 8-6 = 2 & 2 & \(y := z.b.z = aba\) & true\\
    3 & 9-3 = 6 & 6-3 = 3 & 2 & \(w := y.y = abaaba\) & true\\
    0 & 9-0 = 9 & 3-0 = 3 & 1 & \(v := w{\oplus_3}w= (aba)^{3}\) & true\\
  \end{tabular}
\end{center}

Here is the trace of Algorithm~\ref{algo:bin:real} for length \(n := 10\), and \(P := \{ 0,3,6,8 \}\), which is not a period set for $n=10$, i.e., $P$ does not belong to $\Gamma_{10}$.

\begin{center}
  \begin{tabular}{>{\columncolor[gray]{.8}}cccc>{\columncolor[gray]{.8}}ll}
    \rowcolor[gray]{.8}
    period & length & inner period & case & suffix & valid\\
    \hline
    8 & 10-8 = 2 & 10-8= 2 & 2 & \(z := ab\) & true\\
    6 & 10-6 = 4 & 8-6 = 2 & 2 & \(y := z.z = abab\) & true\\
    3 & 10-3 = 7 & 6-3 = 3 & 1 & \(w := y{\oplus_3}y\) & false\\
  \end{tabular}
\end{center}

The table below illustrates that the merge attempted at the last loop iteration for $P[i] = 3$ is impossible, since a mismatch occurs in the overlap.
\begin{center}
\begin{tabular}{>{\columncolor[gray]{.8}}cccccccc>{\columncolor[gray]{.8}}c} 
  \rowcolor[gray]{.8}
  pos. &\color{blue}{0} & 1 & 2 & \color{blue}{3} & 4 & 5 & 6 & 
  \\ 
  $y$ &\texttt{a} &\texttt{b} &\texttt{a} &\color{red}{\texttt{b}} & - & - & - &
  \\ \hline
  $y$ & -         & -         & -         &\color{red}{\texttt{a}}  & \texttt{b} & \texttt{a} & \texttt{b}  &
  \\ \hline
\end{tabular}
\end{center}


\section{Fate and dynamics of period sets}
\label{sec:org219bda4}
\label{sec:fate}

The incremental algorithm and Lemma~\ref{lem:gnm:gn} induces a \emph{parental} relationship between sets in $\Gnm$ and $\Gn$. Any period set $P$ occurs first in \(\Gamma_{max(P)+1}\).  When the length increases from \(n-1\) to \(n\), a period set $P$ in \(\Gnm\) faces three cases: (1) \(P\) may remain as is in $\Gn$, (2) \(P\) has an \emph{extension} with period \(n-1\) (i.e., $P \cup \{n-1\} \in \Gn$), or (3) \(P\) \emph{dies}, i.e., is neither in case (1) nor case (2).  Note that cases (1) and (2) are not exclusive from each other. Thus, $P$ at length $n-1$ can be the parent of at most two period sets in in $\Gn$. One can thus investigate the dynamics of period sets when the word length $n$ increases starting with $n=1$. 

\begin{example}
For instance, $\{0,3,6\}$ is born in $\Gamma_7$ and still belongs to $\Gamma_8$ and $\Gamma_9$; its extension $\{0,3,6,7\}$ also belongs to $\Gamma_8$. From a dynamic view point, $\{0,3,6\}$ is the \emph{parent} of both $\{0,3,6\}$ and $\{0,3,6,7\}$ in $\Gamma_8$.  On the contrary, $\{0,4,6\}$ belongs $\Gamma_7$, but dies at $n=8$, since the pair $(4,6)$ violates the FW condition at $n=8$ (which would require to add $gcd(4,6)$ as a new period). Last, $\{0,2,4,6\}$ belongs to $\Gamma_7$, $\Gamma_8$, and generates $\{0,2,4,6,8\}$ in $\Gamma_9$ because the extension with period $8$ is required by the FPR, but it never dies.
\end{example}

With these definitions at hand, consider the parental relation when the word length $n$ increases starting from $\Gamma_1 := \{ \{ 0 \} \}$: it forms an infinite directed tree whose nodes are period sets and arcs represent the parental relation. The tree is rooted with period set $\{ 0 \}$, is structured in successive layers corresponding to PS for successive word lengths, and the outdegree of each node can be 0, 1, or 2. When a period set dies, a branch of the tree becomes a dead end.

For each period set $P$, one may ask at how many consecutive word lengths it exists. We show that its fate depends only on the periods in $P$, and define below two variables that give the limit of its existence, and give algorithms to compute these.
\begin{definition}[Recursive FW limit and next extension]\label{del:e:rfw}
  Let $P := \{0 < p_1 < \ldots < p_k\}$ with $P \in \Gamma_{max(P)+1}$. 
  The recursive FW limit of $P$, denoted $\rfw(P)$, is defined by  $\rfw(P) := +\infty$ if $p_1 | p_i$ for all $0<i\leq k$, and $\rfw(P) := min_{0<i<j\leq k}\{2p_i-p_j : p_i, p_j \in P \text{ and } 2p_i-p_j > p_k+1\}$. 
  The next extension of $P$, denoted $e(P)$, is  $e(P) := min_{0<i<j\leq k}\{FW(p_i,p_j) : p_i, p_j \in P \text{ and } \gcd(p_i,p_j) \not\in P\}$ if $\card{P}>1$, and $e(\{ 0 \}) = +\infty$ otherwise.
\end{definition}

If $\card{P}>1$, as the word length increases, the current periods of $P$ will induce, by the FPR, new compulsory periods larger than $max(P)$; the minimum among those is $e(P)$, meaning that when the word length reaches $e(P)+1$ then $P$ will necessarily be extended (i.e., possiblity (2) but not (1)), unless $P$ dies.  When the word length reaches $\rfw(P)$, then at least one pair of periods will violate the FW condition, and thus $P$ must die at that length. Of course, if $\card{P} = 1$ or all non zero periods in $P$ are multiple of its basic period, then $\rfw(P) = +\infty$. Thus, any given period set $P$ ceases to belong to $\Gamma_{l}$ if $l := min( e(P)+1, \rfw(P))$ (i.e., case (1) is forbidden), it dies at length $\rfw(P)$, and it is extended at length $e(P)+1$  if $e(P)+1 < \rfw(P)$.

A challenging open question is to compute how many sets dies at length $n$, without enumerating $\Gn$.
The sequence of the numbers of dying period sets at length $n$ is $0, 0, 0, 0, 0, 1, 1, 2, 1, 3, 2, 8$ for $n:= [1,12]$.

\section{Conclusion and exploration of \(\Gn\): distributions of period sets with respect to basic period and to weight}
\label{sec:explore}

\begin{figure}[htbp]
  \centering
  \includegraphics[width=0.9987\textwidth]{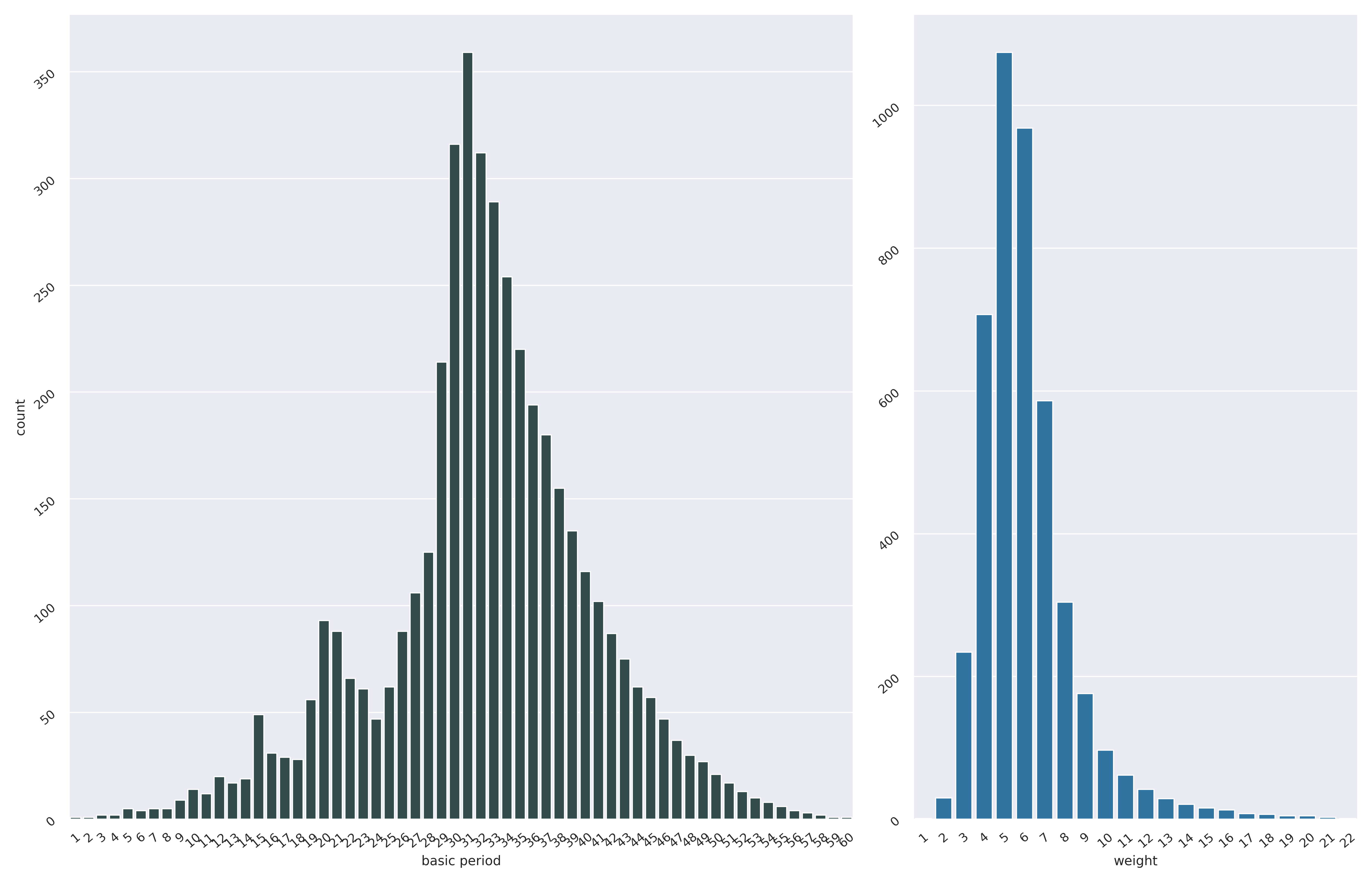}
  \caption{\label{fig:org1e97bb7}\label{fig:basic:knp}Distribution in \(\Gamma_{60}\) of the number of period sets by basic period (left) and by weight (right), for string length of \(n := 60\). Beyond basic period \(30\), the counts decrease smoothly with the basic period. Between basic period \(1\) and \(30\) the counts increase to a local maximum when the basic period reaches $\lfloor n/x \rfloor$ for $1 < x \leq 12 =$ (e.g. basic periods 10, 12, 15, 20, 30). The distribution by weight (right) is limited to weight below $22$; it is unimodal and right skewed towards low weights.}
\end{figure}

The key element of a period set is its basic period, which defines the first level of periodicity in a word. How period sets in \(\Gn\) are distributed according to their basic period is non trivial. Enumerating \(\Gn\) allows inspecting this distribution. The left plot in Figure~\ref{fig:basic:knp} displays \(\Knp\), the counts of period sets for all possible basic periods \(p\), in \(\Gamma_{60}\).
In predicate \(\Xi\)~\cite{GuOd81}, one separates period sets depending on the basic period being \(\leq \lfloor n/2 \rfloor\) (case \textbf{a}) or larger than \(\lfloor n/2 \rfloor\) (case \textbf{b}). The smooth decrease of counts beyond  \(\lfloor n/2 \rfloor\) is explained by the combinatorial property that links number of period sets in case \textbf{b} and the number of binary partitions of an integer (see Lemma 5.8 in~\cite{rivals_jcta_2003}). However, the distribution of counts for period sets in case \textbf{a}, still requires some investigation and statistical modeling. Here, we observe that between basic period \(1\) and \(30\), \(\Knp\) reaches local maxima when \(p\) divides the string length $n$ (e.g. see the peaks at $p = 10, 12, 15, 20$, or $30$, which all correspond to period sets of case \textbf{a}).

Other works have investigated combinatorial parameters that control the number of periods of a word~\cite{gabric_etal_inequality_periods_2021}. Thanks to enumeration of \(\Gn\), one can study the distribution of period sets with respect to weight and how it evolves with \(n\). The right plot of Figure~\ref{fig:basic:knp} displays the number of period sets with equal weight (i.e., same number of periods) for \(n=60\). This distribution is right skewed and illustrates the constraints imposed by multiple periods. Similar figures to Fig.~\ref{fig:basic:knp} for other string lengths are shown in Appendix~\ref{app:distri}.

\textbf{Conclusion}. We provide algorithms to enumerate $\Gn$ incrementally with low space requirement, with several certification functions, and an algorithm for binary realization of a period set. We define a parental relation between period sets for distinct word lengths (which makes up a tree), and propose a way to study their dynamics as $n$ increases. Many questions remain open -- besides that on the number of dying PS mentioned in Section~\ref{sec:fate}. Can one speed up enumeration by exploiting the tree and a dynamic update of the recursive FW limit and next extension of a PS? Can we exploit the tree to unravel how population sizes of PS evolve with $n$? 
Extending the notions presented here to generalizations of words, like partial words, degenerate strings, or multidimensional words opens avenues of future work.  As seen in Section~\ref{sec:fate}, the number of dying PS is not monotonically increasing in function of $n$; thus understanding the sequences of $\Kn$ and $\Knp$ is both stimulating and challenging (see also Figures~\ref{fig:basic:knp}, \ref{fig:distri:qh}-\ref{fig:distri:cn}).

\bigskip\noindent
\textbf{Resource}: we provide files containing the period sets of $\Gn$ for $n = 1, \ldots, 100$ at~\cite{rivals-gamma-data}.

\medskip\noindent
\textbf{Acknowledgements}: This work is part of a project that has received funding from the European Union’s Horizon 2020 research and innovation programme under the Marie Skłodowska-Curie grant agreement No 956229. We thank the anonymous reviewers for their suggestions and comments.

\bibliography{BioJournalDef,combinatorics, rng, eric_rivals_journal, eric_rivals_conf, eric_rivals_preprint, pattern, infobio, DS_bioinfo, eric_rivals_datasets} 

\begin{thebibliography}{10}

\bibitem{Bajic}
Dragana Bajic and Tatjana Loncar-Turukalo.
\newblock A simple suboptimal construction of cross-bifix-free codes.
\newblock {\em Cryptography and Communications}, 6(6):27--37, 2014.
\newblock \href{https://doi.org/10.1007/s12095-013-0088-8}{\path{doi:10.1007/s12095-013-0088-8}}.

\bibitem{Bilotta}
Stefano Bilotta, Elisa Pergola, and Renzo Pinzani.
\newblock A new approach to cross-bifix-free sets.
\newblock {\em IEEE Transactions on Information Theory}, 58(6):4058--4063,
  2012.
\newblock \href {https://doi.org/10.1109/TIT.2012.2189479}
  {\path{doi:10.1109/TIT.2012.2189479}}.

\bibitem{Castelli_1999}
M.Gabriella Castelli, Filippo Mignosi, and Antonio Restivo.
\newblock Fine and wilf’s theorem for three periods and a generalization of
  sturmian words.
\newblock {\em Theoretical Computer Science}, 218(1):83–94, April 1999.
\newblock \href
  {https://doi.org/10.1016/s0304-3975(98)00251-5}
  {\path{doi:10.1016/s0304-3975(98)00251-5}}.

\bibitem{Ehrenfeucht_1979}
Andrzej Ehrenfeucht and D.M. Silberger.
\newblock Periodicity and unbordered segments of words.
\newblock {\em Discrete Mathematics}, 26(2):101–109, 1979.
\newblock \href{https://doi.org/10.1016/0012-365x(79)90116-x}
  {\path{doi:10.1016/0012-365x(79)90116-x}}.

\bibitem{FW65}
Nathan~J. Fine and Herbert~S. Wilf.
\newblock Uniqueness theorems for periodic functions.
\newblock {\em Proc. Amer. Math. Soc.}, 16:109--114, 1965.

\bibitem{gabric_etal_inequality_periods_2021}
Daniel Gabric, Narad Rampersad, and Jeffrey Shallit.
\newblock An inequality for the number of periods in a word.
\newblock {\em International Journal of Foundations of Computer Science},
  32(05):597–614, Jun 2021.
\newblock \href {https://doi.org/10.1142/s0129054121410094}
  {\path{doi:10.1142/s0129054121410094}}.

\bibitem{GuOd81}
Leo~J. Guibas and Andrew~M. Odlyzko.
\newblock Periods in strings.
\newblock {\em J. of Combinatorial Theory series A}, 30:19--42, 1981.
\newblock \href {https://doi.org/10.1016/0097-3165(81)90038-8}
  {\path{doi:10.1016/0097-3165(81)90038-8}}.

\bibitem{Halava:etal:2000}
Vesa Halava, Tero Harju, and Lucian Ilie.
\newblock Periods and binary words.
\newblock {\em J. Comb. Theory, Ser. {A}}, 89(2):298--303, 2000.
\newblock \href{https://doi.org/10.1006/JCTA.1999.3014} {\path{doi:10.1006/JCTA.1999.3014}}.

\bibitem{holub_shallit_random_words}
Stepan Holub and Jeffrey~O. Shallit.
\newblock Periods and borders of random words.
\newblock In Nicolas Ollinger and Heribert Vollmer, editors, {\em 33rd
  Symposium on Theoretical Aspects of Computer Science, {STACS} 2016, February
  17-20, 2016, Orl{\'{e}}ans, France}, volume~47 of {\em LIPIcs}, pages
  44:1--44:10. Schloss Dagstuhl - Leibniz-Zentrum f{\"{u}}r Informatik, 2016.
\newblock \href {https://doi.org/10.4230/LIPIcs.STACS.2016.44}{\path{doi:10.4230/LIPIcs.STACS.2016.44}}.

\bibitem{Knuth-Morris-Pratt-pm}
Donald~E. Knuth, James~H. Morris, and Vaughan~R. Pratt.
\newblock Fast pattern matching in strings.
\newblock {\em SIAM Journal of Computing}, 6:323--350, 1977.
\newblock \href {https://doi.org/10.1137/0206024} {\path{doi:10.1137/0206024}}.

\bibitem{Leopardi:2009}
Paul Leopardi.
\newblock Testing the {Tests}: {Using} {Random} {Number} {Generators} to
  {Improve} {Empirical} {Tests}.
\newblock In Pierre~L' Ecuyer and Art~B. Owen, editors, {\em Monte {Carlo} and
  {Quasi}-{Monte} {Carlo} {Methods} 2008}, pages 501--512. Springer Berlin
  Heidelberg, 2009.
\newblock DOI: 10.1007/978-3-642-04107-5\_32.
\newblock URL:
  \url{http://link.springer.com/chapter/10.1007/978-3-642-04107-5_32}.

\bibitem{Lothaire-COW-97}
M.~Lothaire, editor.
\newblock {\em Combinatorics on Words}.
\newblock Cambridge University Press, second edition, 1997.

\bibitem{Lothaire-ACW}
M.~Lothaire.
\newblock {\em Algebraic Combinatorics on Words}.
\newblock Cambridge University Press, Cambridge, 2005.
\newblock URL: \url{http://www-igm.univ-mlv.fr/\string~berstel/Lothaire/index.html}.

\bibitem{MAR-ZAM-1993}
George Marsaglia and Arif Zaman.
\newblock Monkey tests for random number generators.
\newblock {\em Computers and Mathematics with Applications}, 26(9):1--10, 1993.

\bibitem{Nielsen_bifix_1973}
Peter~Tolstrup Nielsen.
\newblock A note on bifix-free sequences (corresp.).
\newblock {\em IEEE Transactions on Information Theory}, 19(5):704–706,
  September 1973.
\newblock \href{https://doi.org/10.1109/tit.1973.1055065}{\path{doi:10.1109/tit.1973.1055065}}.

\bibitem{Nielsen_expect_search_1973}
Peter~Tolstrup Nielsen.
\newblock On the expected duration of a search for a fixed pattern in random
  data (corresp.).
\newblock {\em IEEE Transactions on Information Theory}, 19(5):702–704,
  September 1973.
\newblock \href{https://doi.org/10.1109/tit.1973.1055064}{\path{doi:10.1109/tit.1973.1055064}}.

\bibitem{PER:WHI:1995}
Ora~E. Percus and Paula~A. Whitlock.
\newblock {Theory and Application of Marsaglia's Monkey Test for Pseudorandom
  Number Generators}.
\newblock {\em ACM Transactions on Modeling and Computer Simulation},
  5(2):87--100, April 1995.

\bibitem{C_cpm_common_words_2000}
Sven Rahmann and Eric Rivals.
\newblock Exact and efficient computation of the expected number of missing and
  common words in random texts.
\newblock In {\em Proc. of CPM 2000}, page 375–387. Springer Berlin
  Heidelberg, 2000.
\newblock \href{https://doi.org/10.1007/3-540-45123-4_31}{\path{doi:10.1007/3-540-45123-4_31}}.

\bibitem{rahmann_cpc_2003}
Sven Rahmann and Eric Rivals.
\newblock On the distribution of the number of missing words in random texts.
\newblock {\em Combinatorics, Probability and Computing}, 12(01), Jan 2003.
\newblock \href{https://doi.org/10.1017/s0963548302005473}{\path{doi:10.1017/s0963548302005473}}.

\bibitem{rivals-gamma-data}
Eric Rivals.
\newblock Sets of period sets for words of length n.
\newblock Zenodo, Nov 2024.
\newblock Data set.
\newblock \href{https://doi.org/10.5281/zenodo.13826259}{\path{doi:10.5281/zenodo.13826259}}.

\bibitem{C_icalp_period_2001}
Eric Rivals and Sven Rahmann.
\newblock Combinatorics of periods in strings.
\newblock In {\em Proc. of ICALP 2001}, page 615–626. Springer Berlin
  Heidelberg, 2001.
\newblock \href{https://doi.org/10.1007/3-540-48224-5_51}{\path{doi:10.1007/3-540-48224-5_51}}.

\bibitem{rivals_jcta_2003}
Eric Rivals and Sven Rahmann.
\newblock Combinatorics of periods in strings.
\newblock {\em Journal of Combinatorial Theory, Series A}, 104(1):95--113, Oct
  2003.
\newblock \href{https://doi.org/10.1016/s0097-3165(03)00123-7}{\path{doi:10.1016/s0097-3165(03)00123-7}}.

\bibitem{C_icalp_2023}
Eric Rivals, Michelle Sweering, and Pengfei Wang.
\newblock {Convergence of the Number of Period Sets in Strings}.
\newblock In Kousha Etessami, Uriel Feige, and Gabriele Puppis, editors, {\em
  50th International Colloquium on Automata, Languages, and Programming (ICALP
  2023)}, volume 261 of {\em Leibniz International Proceedings in Informatics
  (LIPIcs)}, pages 100:1--100:14, Dagstuhl, Germany, 2023. Schloss Dagstuhl --
  Leibniz-Zentrum f{\"u}r Informatik.
\newblock \href{https://doi.org/10.4230/LIPIcs.ICALP.2023.100}{\path{doi:10.4230/LIPIcs.ICALP.2023.100}}.

\bibitem{Smyth-book-03}
William~F. Smyth.
\newblock {\em Computating Pattern in Strings}.
\newblock Pearson - Addison Wesley, 2003.

\bibitem{Tijdeman_2003}
Robert Tijdeman and Luca~Q. Zamboni.
\newblock Fine and wilf words for any periods.
\newblock {\em Indagationes Mathematicae}, 14(1):135–147, March 2003.
\newblock \href{https://doi.org/10.1016/s0019-3577(03)90076-0}{\path{doi:10.1016/s0019-3577(03)90076-0}}.

\bibitem{Tijdeman_2009}
Robert Tijdeman and Luca~Q. Zamboni.
\newblock Fine and wilf words for any periods ii.
\newblock {\em Theoretical Computer Science}, 410(30–32):3027–3034, August
  2009.
\newblock \href{https://doi.org/10.1016/j.tcs.2009.02.004}{\path{doi:10.1016/j.tcs.2009.02.004}}.

\bibitem{Ukkonen-qgram-TCS}
Esko Ukkonen.
\newblock Approximate string-matching with $q$-grams and maximal matches.
\newblock {\em Theor. Comp. Sci.}, 92(1):191--211, January 1992.
\newblock \href{https://doi.org/10.1016/0304-3975(92)90143-4}{\path{doi:10.1016/0304-3975(92)90143-4}}.

\end{thebibliography}

\newpage
\appendix
\renewcommand{\topfraction}{0.84}
\renewcommand{\bottomfraction}{0.6}
\renewcommand{\textfraction}{0.1}
\renewcommand{\floatpagefraction}{0.70095}

\section{Distributions of the number of period sets by basic period and by weight}
\label{sec:distri}\label{app:distri}

\begin{figure}[htbp]
  \centering
  \includegraphics[width=0.9987\textwidth]{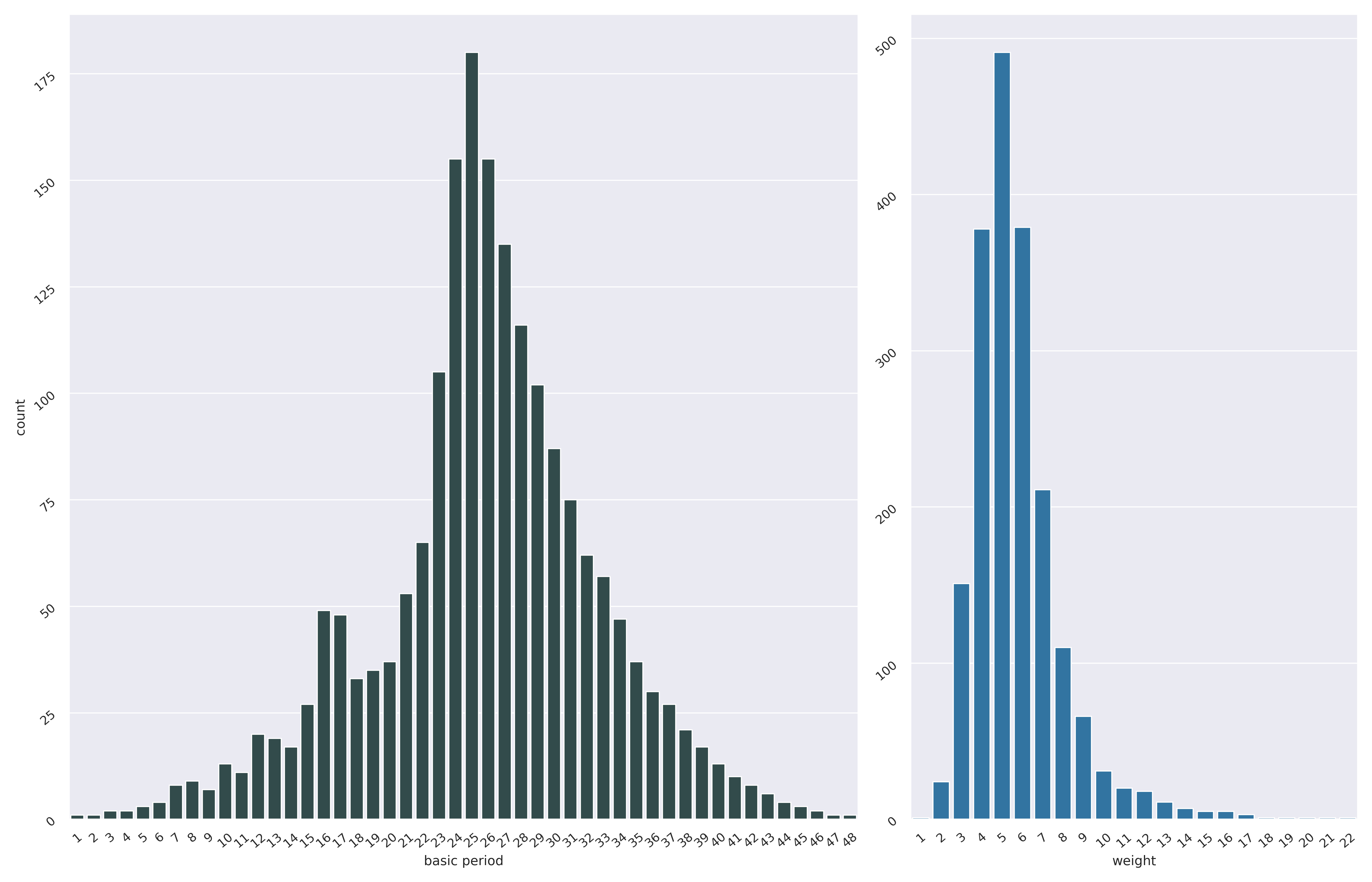}
  \caption{\label{fig:distri:qh}Distribution in \(\Gamma_{48}\) of the number of period sets by basic period (left) and by weight (right), i.e., for string length of \(n := 48\).}
\end{figure}

\begin{figure}[htbp]
  \centering
  \includegraphics[width=0.9987\textwidth]{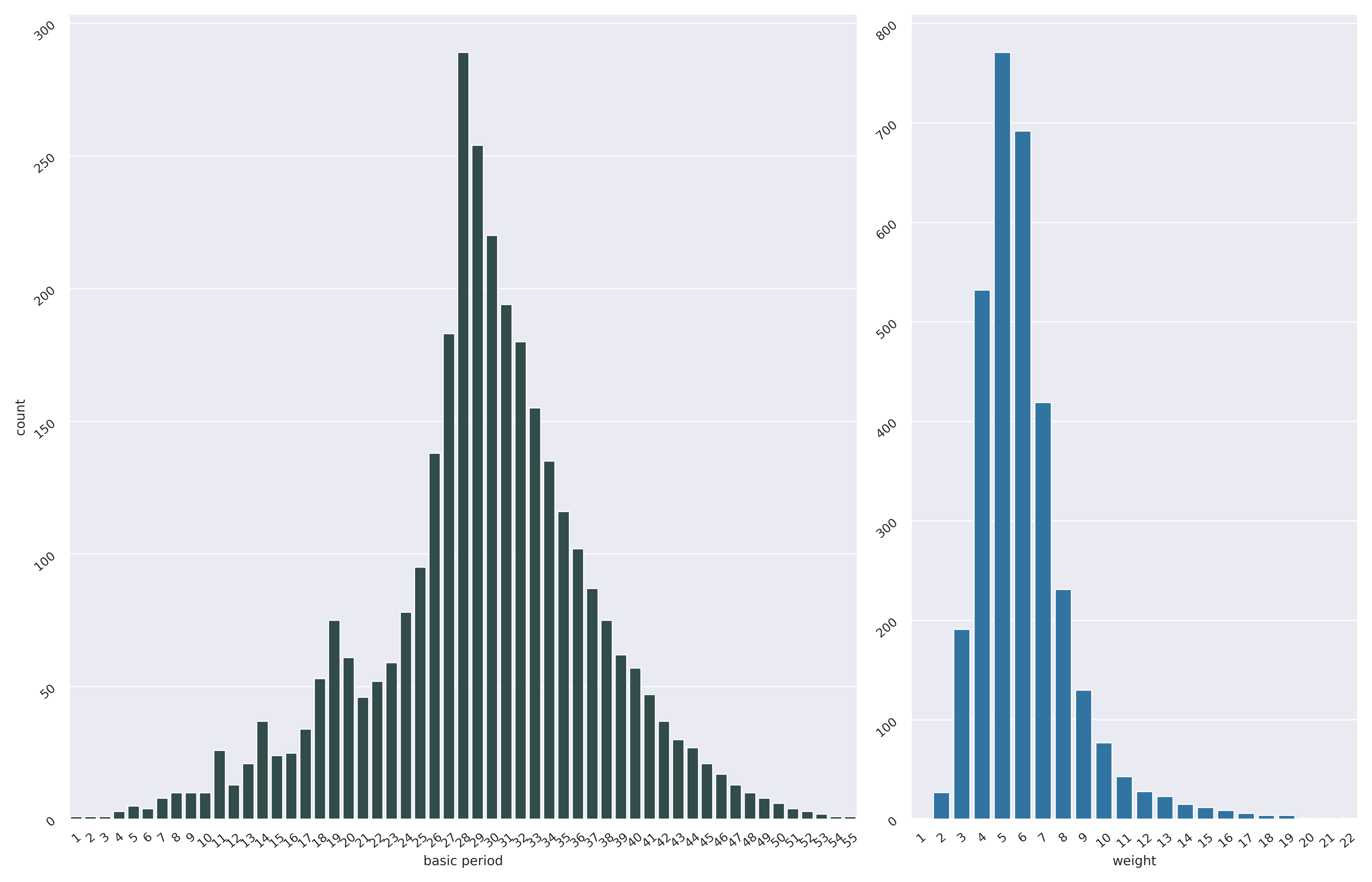}
  \caption{\label{fig:distri:cc}Distribution in \(\Gamma_{55}\) of the number of period sets by basic period (left) and by weight (right), i.e., for string length of \(n := 55\).}
\end{figure}

\begin{figure}[htbp]
  \centering
  \includegraphics[width=0.9987\textwidth]{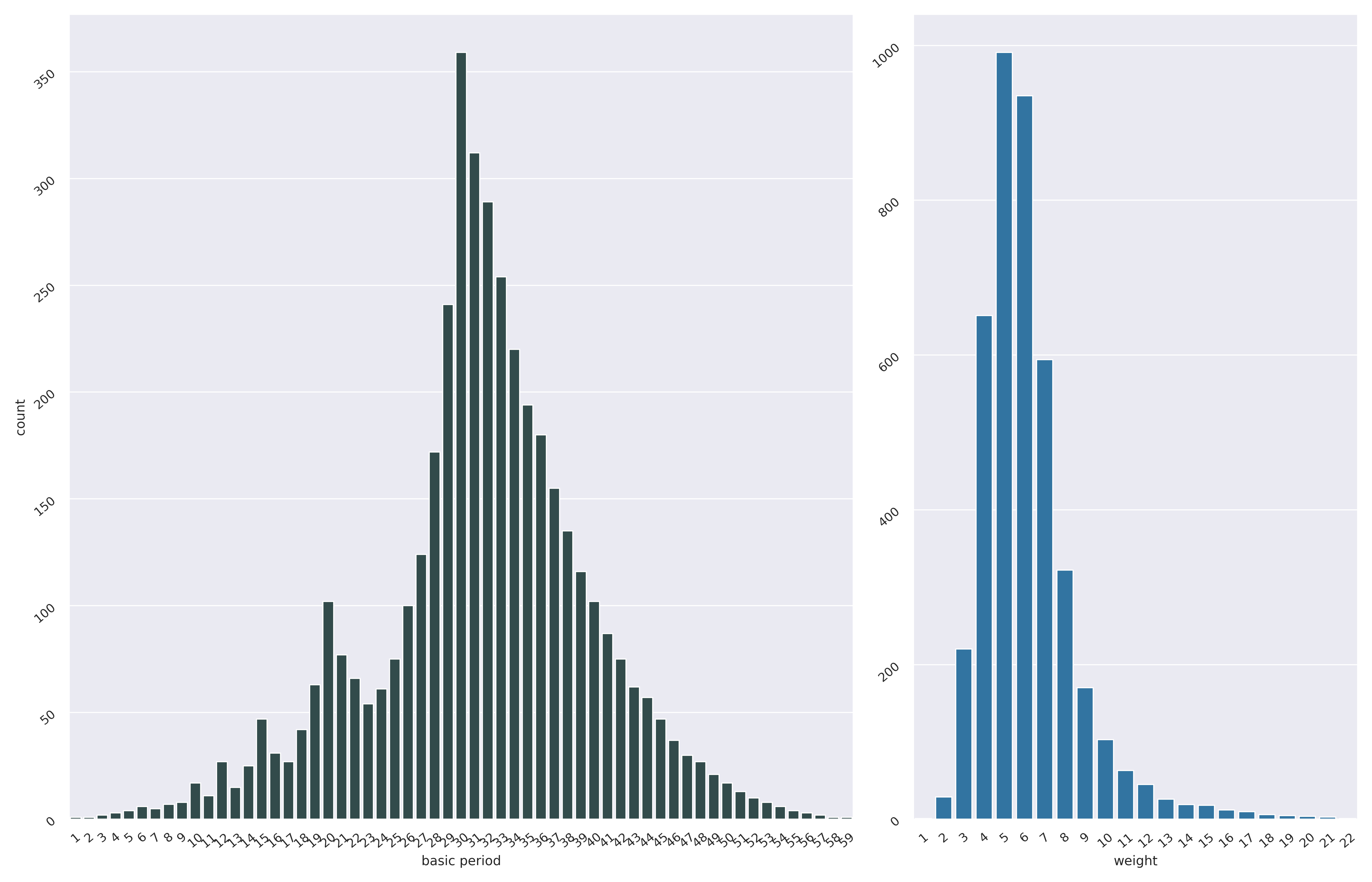}
  \caption{\label{fig:distri:cn}Distribution in \(\Gamma_{59}\) of the number of period sets by basic period (left) and by weight (right), i.e., for string length of \(n := 59\).}
\end{figure}

Like in Figure~\ref{fig:basic:knp}, we explore how period sets are distributed according to their basic period, and according to their weight for other string lengths.
We plot these distributions for $n = 48$, $n = 55$, and $n = 59$ in Figures~\ref{fig:distri:qh},  \ref{fig:distri:cc}, and~\ref{fig:distri:cn},  respectively. We choose these values because they differ in their number of divisors $48 = 2^4 \times 3$, $55 = 5 \times 11$ and $59$ is prime. In essence, both plots for \(\Gamma_{48}\), \(\Gamma_{55}\), and \(\Gamma_{59}\) look very similar to those for \(\Gamma_{60}\). Even for a prime string length, $n=59$, the distribution of number of period sets in case \textbf{a}, shows a maximum at $\lfloor n/2 \rfloor$ and local maxima at $\lfloor n/3 \rfloor$, $\lfloor n/4 \rfloor$ etc. 


\section{Alternative certification functions and variants of incremental algorithm}
\label{sec:app:alternative}

\subsection{Incremental algorithm with rule based certification}
\label{sec:inc:rule}
Here, we detail an alternative version of the incremental algorithm, which uses the rule based certification function derived from Theorem 5.1 from~\cite{GuOd81} (see also below). This is related to subsection~\ref{sec:alt}. Algorithm~\ref{algo:inc} presents the pseudo-code; it uses two functions named \emph{checkFPR} and \emph{checkBPR}, which check if a set of integers satisfies respectively, the Forward and Backward Propagation Rules.

In our case, as the candidate sets include a period set of \(\Gnm\), they necessarily satisfy the first condition (i).
Regarding the FPR, since \(P\) belongs to \(\Gnm\),  \(P\) satisfies the FPR up to position \(n-2\) included; and thus, only period \(n-1\) can be required by the FPR. For each possible pair \((p,q)\) considered in the FPR, we only need to check if the FPR formula yields \((n-1)\). Second, for the same reason, when considering the candidate set \(P \cup \{n-1\}\), we are sure that the FPR is satisfied.

Let \(R\) be any subset of \(\{0, 1, \ldots, n-1\}\) containing zero and assuming that \(R\) is sorted in increasing order, then
we have that checking the FPR and BPR takes \(O(n\log(n))\) time.

Algorithm~\ref{algo:inc} differs from Algorithm~\ref{algo:inc:gen} in two aspects. First, it can indicate for which reason the candidate set is not a period set if the check fails. Second, it also computes the set of "dying" period sets of \(\Gnm\), that is the period set of \(\Gnm\) that are not in  \(\Gn\), nor cannot be extended at length \(n\).
We will define these notions in Section~\ref{sec:fate}. Of course, dying period sets could also be computed within Algorithm~\ref{algo:inc:gen}, which uses the predicate \(\Xi\) (but for simplicity and to avoid redundancy, was not mentioned earlier).

Altogether the time complexity of Algorithm~\ref{algo:inc} is bounded by \(O(n\log(n) \times \Kn)\), which may not be optimal.

\begin{definition}\label{def:dying}
A dying period set \(P\) is a period set of \(\Gnm\) such that neither \(P\) nor \(P \cup \{ n-1 \}\) belong to \(\Gn\). 
\end{definition}

\begin{algorithm}[H]
  \SetKwInOut{Input}{Input}
  \caption{\label{algo:inc} IncrementalGamma with rule based certification}
  \Input{$n>1$: integer; $\Gamma_{n-1}$: the set of period sets for length $n-1$} 
  \textbf{Output}: {$\Gamma_n$: the set of period sets for length $n$; $D$: the set of dying PS at length $n$}\;
  \BlankLine
  \nl $G := \emptyset$\tcp*{   $G$: variable to store $\Gamma_n$}
  \nl $D := \emptyset$\tcp*{   $D$: variable to store dying PS}
  \nl \For{ $P \in \Gamma_{n-1}$ }{
    \nl  $Q := P \cup \{ n-1 \}$ \tcp*[f]{build extension of $P$ with period $n-1$}\;
    \nl \If(\tcp*[h]{$n-1$ is required by FPR at length $n$}){checkFPR( $P$, $n$ )}{
      \nl \lIf{ checkBPR( $Q$, $n$ ) }{ insert $Q$ in $G$}
      \nl \lElse{ insert $P$ in $D$ \tcp*[f]{otherwise $P$ is dying at length $n$}}
      }
     \nl\Else{
         \nl \lIf{ checkBPR( $P$, $n$ ) }{ insert $P$ in $G$ }
         \nl \Else{
           \nl \lIf{ checkBPR( $Q$, $n$ ) }{ insert $Q$ in $G$ }
           \nl \lElse{ insert $P$ in $D$ \tcp*[f]{otherwise $P$ is dying at length $n$}}
         }
       }
     }
    \nl   \KwRet{$G$ and $D$}\;
\end{algorithm}

\subsection{An alternative combinatorial certification}
\label{sec:alt:acw}

A third certification function exploits the characterization of Theorem~\ref{th:acw:charac}, where the necessary conditions are expressed in function of the periods $p_h$, their differences $d_h$, for $0 < h \leq s$, and of $n$. In the incremental approach, the word size increases from $n-1$ to $n$. Here we sketch why the certifications of candidate sets $P$ and $Q := P \cup \{n-1\}$ can be combined in $O(n)$ time.

First, from candidate set $P$ containing $s+1$, we can check whether period $n-1$ can be deduced from smaller periods: if there is an $h$ such that $p_h+d_h = n-1$. This can be done in $s$ constant time computations. If yes, then $n-1$ is a compulsory period at length $n$ and thus only $Q$ must be further examined. If not, both $P$ and $Q$ may belong to $\Gn$. Doing the above computations, we determine for which indices $h$ the condition  $d_h + p_h = n-1$ is satisfied, and only for those we have to check conditions (a) and (b). Indeed, for all $h$ such that $d_h + p_h < n-1$, we know (a) and (b) are already satisfied since $P \in \Gnm$.
However, at length $n$, the period $p_s$ of $P$ becomes $n$ which changes the value $d_s$ from $n-1$ to $n$. Hence, we must check condition (b), only for $h = s-1$, which is done in constant time.

In the case of $Q$, the number of periods is now $s+2$, the new period is $p_s=n-1$, $d_s$ has a new value, and $d_{s+1} = 1$, since the trivial period $n$ is included in $Q$. The condition (a) is then always satisfied. The condition (b) needs to be verified only for $h$ equal to $s-1$ and $s$, since only  $d_s$ and $d_{s+1}$ have changed. This can be done in two constant time computations.
Altogether, with this combined certifications of $P$ and $Q$, the inner for loop of Algorithm~\ref{algo:inc:gen} takes $O(s)$ time, which we can bound by $O(n)$.

\section{Algorithm Binary realization}
\label{sec:bin:real:ten}

\subsection{Correctness and complexity of the algorithm}
\label{app:correct}

\begin{proof}
Let us prove that the Algorithm Binary Realization is correct.

\textbf{Correction of the base case}
As we process the last period of \(P\), the nested set is \(\{0\}\) for length \(n-max(P)\). We must build a suffix without period (i.e., whose basic period is its length). Hence, the word  \(\mathtt{a.b^{(\text{prevLg} -1)}}\) is a binary realization for this set.

\textbf{Correction of the general case.}
After setting variables \emph{lg} and \emph{innerPeriod}, we check the condition (\emph{innerPeriod} \(<\) \emph{prevIP}). In a period set, the offset \(P[i+1]-P[i]\) decreases when \(i\) increases.  The condition implies the current nested set is invalid, and we return \(\epsilon\) as needed. Another way to formulate this: If the condition is satisfied, then \emph{suffix}, which ends with \emph{prevSuffix}, does not satisfy the FPR, meaning that this set is invalid.

The invariant at the start of the for loop is that \emph{prevSuffix} realizes the nested set \(P_{P[i+1]}\) and has \emph{prevIP} as basic period. By construction, we know that \emph{lg} \(=\) \emph{prevLg} \(+\) \emph{innerPeriod}. By construction, \emph{suffix} ends with \emph{prevSuffix} and has basic period \emph{innerPeriod}. Thus, by the invariant, \emph{suffix} will realize \(P_{P[i]}\).

\textbf{Case 1}
We build \emph{suffix} by concatenating a prefix of \emph{prevSuffix} of length \emph{innerPeriod} with \emph{prevSuffix} (line 10), and we must ensure that \emph{suffix} has basic period \emph{innerPeriod}.
Let us consider the conditions from line 9.
\begin{enumerate}
\item If ( \emph{innerPeriod} \(=\) \emph{prevIP} ) then, as \emph{prevSuffix} already has period \emph{prevIP}, \emph{suffix} will inherit from it.
Otherwise we know that ( \emph{innerPeriod} \(>\) \emph{prevIP} ).
\item Then, \emph{prevSuffix} has a basic period (\emph{prevIP}) that should not divide \emph{innerPeriod}, which is the length of the prefix of \emph{prevSuffix} that occurs as prefix of \emph{suffix}. Hence, we require the condition (\emph{prevIP} \(\nmid\) \emph{innerPeriod}) to be satisfied. Otherwise, \emph{suffix} would also have \emph{prevIP} as period;  then \emph{suffix} would be a binary world, but would not realize \(P\).
\item Then, if (\emph{innerPeriod} \(=\) \emph{prevLg}) then \emph{lg} \(= 2\times\) \emph{prevLg} and \emph{suffix} equals \(\text{/prevSuffix/}^2\) and has the desired length and basic period.
\item Otherwise, we check that \emph{prevSuffix} has period \emph{innerPeriod}. If yes, then \emph{suffix} also has period \emph{innerPeriod} by construction (line 10), and thus realizes \(P_{P[i]}\).
If not, then there is no possible realization of \(P\) and we return \(\epsilon\) (line 11).
\end{enumerate}

\textbf{Case 2}
Here, we know that \emph{lg} is larger than twice \emph{prevLg}. Therefore, we will build a prefix that starts with \emph{prevSuffix} followed by \emph{m} new symbols, such that \emph{suffix} has no period shorter than \emph{innerPeriod}. Hence, we must ensure that \emph{newPrefix} is primitive, otherwise it would have a period that divides \emph{innerPeriod}. By Lemma 3 from~\cite{Halava:etal:2000}, for any binary word \(w\), \(w\mathtt{a}\) or \(w\mathtt{b}\) is primitive. So, we concatenate \(a^{m}\) to \emph{prevSuffix}, and check if it is primitive (in \(O(\length{\text{newPrefix}})\) time). If not, we change its last symbol $\mathtt{a}$ by $\mathtt{b}$. In both cases, \emph{newPrefix} is primitive. By construction, \emph{suffix} has basic period \emph{innerPeriod} as desired, and thus realizes \(P_{P[i]}\).
\end{proof}






\section{Checking FPR and BPR}
\label{sec:fpr}

Let us state some properties:
\begin{enumerate}
\item From the definition of FPR, we can see that checking the FPR for a pair \((p,q)\) of \(P\) is equivalent to checking the FPR for pair \((0,q)\) in the nested PS \(P_p\).
\item Assume the FPR is satisfied for pair \((0,p)\). Then, it is also satisfied for any pair \((hp, jp)\) with \(1\leq h < j <\lfloor n/p \rfloor\) and \(hp\),\(jp \in P\), since both periods are multiples of \(p\).
\end{enumerate}

From both properties, we get that once the FPR has been checked for the first pair \((p,q)\) taken that has offset \((q-p)\), it is also satisfied for any other pair whose offset equals \(r\) or a multiple of \(r\). It follows that, for a set \(P\), one can limit the checking of FPR only to left most pairs whose offsets differ from eachother and are not multiple of another offset. Thus, at least one element, say \(p\), must be an \emph{irreducible period} (as defined in~\cite{rivals_jcta_2003}), and \(q\) is the closest period to \(p\) (i.e., one which gives rise to the smallest offset with respect to \(p\)). Since, the number of irreducible periods of a period set of  \(\Gn\) is bounded by \(\log_2(n)\)~\cite{C_icalp_2023}, the number of such pairs also is. We obtain the bound on the complexity for the general case stated in Lemma~\ref{lem:fpr:comp}.

\subsection{Checking the FPR and the BPR}
\label{sec:orgf92c0dc}
Let \(n>0\) and \(P\) be a subset of \(\{0, 1, \ldots, n-1\}\). We assume that \(P\) is given as an ordered array. The complexity for checking the FPR or the BPR for \(P\),  has, to our knowledge not been previously addressed. For any pair \(p < q\), we call their difference \((q-p)\), an offset.

\textbf{Checking the BPR}.
Here, we demonstrate a property that relates the BPR to the FW Theorem. Precisely, if BPR is violated for some pair \((p,q)\) at length \(n\), with period \(r := p-i(q-p)\) for some \(i\), then the pair \((p-r, q-r)\) violates the FW condition of Theorem~\ref{thm:gcd:cond} in the nested set of length \((n-r)\).
\begin{lemma}\label{lem:bpr:fw}
  Let \((p,q)\) be a pair of integers that violate the BPR, and let $i\geq 2$ such that \(r := p-i(q-p) \in P\). Then the pair \((p-r, q-r)\) violates the FW condition for length \((n-r)\).
\end{lemma}

\begin{proof}
  Let \(P \in \Gn\). Let \(p,q\) in \(P\) satisfying \(0\leq p < q < 2p\) be such that \((2p-q) \not\in P\). Assume \((p,q)\) violates the BPR. Then, there exists \(i\) in \([2,\ldots,\min(\lfloor p/(q-p), \rfloor, \lfloor (n-p)/(q-p) \rfloor)]\), such that \(p-i(q-p) \not \in P\). If several such integers exist, choose \(i\) as their minimum, and define \(r := p-i(q-p)\). We will show that the nested period set of \(P\) for length \((n-r)\) is not a period set, since two of its periods violates the FW condition, which would require their gcd as an additional period, thereby implying that \(P \not \in \Gn\), a contradiction. Since \(i\) is chosen minimal, we have that \(\gcd(p,q)\) is not in \(P\) by the definition of the BPR.
  Note that \(p-r = i(q-p)\) and \(q-r = (i+1)(q-p)\). Thus, one gets \(gcd(p-r, q-r) = q-p\), and the FW limit of \((p-r, q-r)\) equals \(2i(q-p)\). Indeed, \(FW(p-r, q-r) := p-r + q-r - gcd(p-r, q-r) = 2p-2r = 2i(q-p)\).
  By hypothesis, we have:
  \[
    \begin{array}{llcl}
      ~
                      &i            &\leq &(n-p)/(q-p) \\
      \Leftrightarrow &p+i(q-p)     &\leq &n \\
      \Leftrightarrow &r+2i(q-p)    &\leq &n \\
      \Leftrightarrow &FW(p-r, q-r) &\leq &n-r
    \end{array}
  \]
   meaning that \((p-r, q-r)\) violates the FW condition of Theorem~\ref{thm:gcd:cond} for length \((n-r)\). 
\end{proof}
In algorithmic terms, checking the BPR of a set $P$ can be done by checking the FW condition of Theorem~\ref{thm:gcd:cond} in each nested set of $P$. Altogether this takes $O(\card{P})$ time and space.

\textbf{Checking the FPR}. Some properties are explained in Appendix~\ref{sec:fpr} and lead to this Lemma.
\begin{lemma}\label{lem:fpr:comp}
Let \(P\) is a subset of \([0, \ldots, n-1]\), that is ordered, and has zero as first period. Checking the FPR for \(P\) in general takes \(O(n\log(n)) time\).
\end{lemma}

\section{Fate: computation of the limits of a period set}
\label{sec:sup:fate}

\subsection{Next extension}
\label{sec:org219f9c7}

Algorithm~\ref{algo:ext:limit} computes the next extension of \(P\). The next extension is a length at which some deducible period needs to be added to \(P\) to satisfy the FPR. It equals the added period plus one, and must be larger than the birth length of \(P\) (Indeed, \(P \in \Gamma_{max(P)+1}\), and thus satisfies the FPR for that length). By definition of the FPR, a period induced by the FPR equals \(P[i] + P[i]-P[j]\) for some indexes \(0 < j < i < \card{P}\). Because, we need the minimum of added periods, we can restrict the computation to pairs of adjacent periods (i.e., that is to case where \(j=i-1\)), since the offset between periods decreases with their index. Hence, the formula \(P[i] + (P[i]-P[i-1])\) for computing the limit induced from current period \(P[i]\). Because of this, we can also rule out cases where \(P[i]\) is smaller the half the birth length of \(P\) (line 6).

\begin{algorithm}[H]
  \SetKwInOut{Input}{Input}
  \caption{\label{algo:ext:limit} next extension( period set $P$  (in a sorted array))} 
  \textbf{Output}: {$e(P)$}\;
  \BlankLine
  \nl $\text{birthLg} := max(P)+1$\tcp*{min $x$ s.t. $P$ belongs to $\Gamma(x)$}
  \nl $\text{limit} := +\infty$    \tcp*{limit to be computed, init. with $+\infty$}
  \nl \For{$i := \card{P}-1$ to $1$ }{
    \nl \If(\tcp*[h]{ }){$P[i] \leq \lfloor \frac{\text{birthLg}}{2} \rfloor$ }{
      \nl break\tcp*{avoid such $P[i]$ values whose limit cannot be $> \text{birthLg}$}
    }
    \nl \If(\tcp*[h]{ current limit is beyond $\text{birthLg}$}){$P[i] + (P[i]-P[i-1]) \geq \text{birthLg}$ )}{
      \tcp{update $\text{limit}$ with the min of $\text{limit}$ and current limit}
      \nl $\text{limit} := min( \text{limit}, P[i] + (P[i]-P[i-1]) )$\;
    }
  }
  \nl   \KwRet{$\text{limit}$}\;
\end{algorithm}
\subsection{Recursive FW limit}
\label{sec:org1b88413}
We exhibit an algorithm to compute what we termed, the recursive FW limit of a PS \(P\) (see Algorithm~\ref{algo:rec:fw:limit}). The FW Theorem provides a way to compute a maximal length for any pair of distinct, non trivial periods such that one period is not a multiple of the other. For any \(p,q\) in \(P\) such that \(0 < p < q < n\) and \(p \nmid q\), we denote by \(FW(p,q)\) the FW limit, that is
\(FW(p,q) := p+q - gcd(p,q)\). If \(p\div q\) we assume that \(FW(p,q) := +\infty\).
First, the algorithm proceeds with two special cases: if all periods are multiple of the basic period, then it returns \(+\infty\). Note this includes the case with basic period equals to one. If \(P\) contains only three periods, then it returns \(FW(P[1],P[2])\).

Otherwise, it will compute the limit \(l\) and initializes with \(+\infty\).  It loops over \(P\) backwards, to consider longer and longer suffixes starting at a position with period of a word satisfying \(P\), and builds a list \(Q\) of periods restricted to the current suffix. The periods in \(Q\) are those of \(P\) minus the starting position. It computes \(FW(Q[1],Q[2])\) and takes the minimum between \(l\) and \(P[i] + FW(Q[1],Q[2])\). After terminating the loop, it returns the limit \(l\).

\begin{algorithm}[H]
  \SetKwInOut{Input}{Input}
  \caption{\label{algo:rec:fw:limit} RecursiveFWLimit( period set $P$ in a sorted array )} 
  \textbf{Output}: {the minimum length at which a pair of periods of $P$ requires a change of basic period (application of FW Theorem)}\;
  \BlankLine
  \nl \If(\tcp*[h]{If basic period divides all other periods}){$(P[1] \mid P[i])$ for all $1 < i < \card{P}$ }{
    \nl \KwRet{$+\infty$}\;
  }
  \nl \If(\tcp*[h]{If $P$ contains only two non trivial periods}){$\card{P} = 3$}{
    \nl \KwRet{$FW(P[1], P[2])$}\;
  }
  
  \nl $\text{limit} := +\infty$  \tcp*{limit to be computed, init. with $+\infty$}
  \nl insert $(P[n-1] - P[n-2])$ in $Q$ \tcp*{Init $Q$ with the last offset between periods}
  \nl \For{$i := \card{P}-3$ to $0$ }{
    \nl $\text{offset} := P[i+1] - P[i]$\;
    \nl $Q[0] := Q[0] + \text{offset}$\;
    \nl insert $\text{offset}$ at first position in $Q$\;
    \nl $\text{limit} := min( \text{limit}, P[i] + FW(Q[0], Q[1]) )$ \;
  }
  \nl   \KwRet{$\text{limit}$}\;
\end{algorithm}
\textbf{Complexity}. In Algorithm~\ref{algo:rec:fw:limit},  the first special case is processed in \(\card{P}\) time (lines 1--2), while the second one requires constant time (lines 3--4). The main loop is executed at most \(\card{P}\) times and all instructions in it take constant time (lines 7--11). Altogether,  Algorithm~\ref{algo:rec:fw:limit} takes \(O(\card{P})\) time and constant space.

\textbf{Correctness}.  The correctness of Algorithm~\ref{algo:rec:fw:limit} follows from Lemma~\ref{lem:bpr:fw}.

\section{Properties of periods and characterization of period sets} \label{go:charac}


\subsection{Properties of periods}
\label{app:periods}

Let us state some known, useful properties of periods, which are detailed in~\cite{C_icalp_2023}.
\begin{lemma}\label{lem:multiply}
Let $p$ be a period of $u \in \Sigma^n$ and $k \in \mathbb{N}_{\geq 0}$ such that $kp < n$. Then $kp$ is also a period of $u$.
\end{lemma}
\begin{lemma}\label{lem:add}
Let $p$ be a period of $u \in \Sigma^n$ and $q$ a period of the suffix $w = u[p\dd n-1]$. Then $(p + q)$ is a period of $u$. Moreover, $(p + kq)$ is also a period of $u$ for all $k \in \mathbb{N}_{\geq 0}$ with $p + kq < n$.
\end{lemma}
\begin{lemma}\label{lem:subtract}
Let $p, q$ be periods of $u \in \Sigma^n$ with $0 \leq q \leq p$. Then the prefix and the suffix of length $(n-q)$ have the period $(p-q)$.
\end{lemma}

\begin{lemma}\label{lem:divide}
Suppose $p$ is a period of $u \in \Sigma^n$ and there exists a substring $v$ of $u$ of length at least $p$ and with period $r$, where $r|p$. Then $r$ is also a period of $u$.
\end{lemma}

\subsection{Characterization of autocorrelations/period sets \cite{GuOd81}}
\label{sec:xi} \label{sec:orgc181505}
Guibas and Odlyzko have provided two equivalent characterizations of period sets: one is given by predicate $\Xi$, the other is the rule based characterization given in Section~\ref{sec:prelim}. However, they manipulate period sets as binary vectors called \emph{autocorrelation} (or sometimes correlation for short). Remind that an autocorrelation is a binary encoding in a binary string of length $n$ of a period set of $\Gn$. We recall in extenso the original predicate $\Xi$ and then their Theorem 5.1, which states the equivalence of characterizations and the alphabet independence.

\begin{t}
  \centering
  \includegraphics[width=0.87\textwidth]{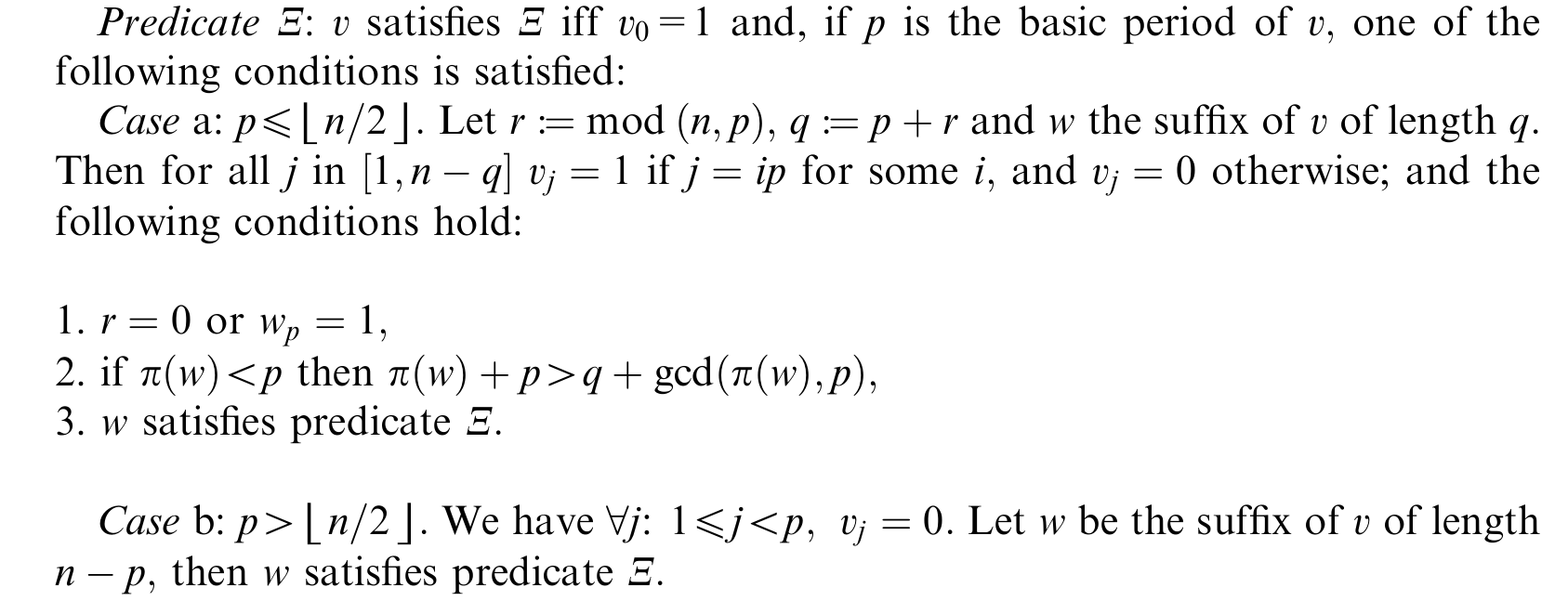}
\end{t}

\begin{theorem}
  Let \(v\) a binary string of length \(n\). The following statements are equivalent:
  \begin{enumerate}
  \item \(v\) is the autocorrelation of a binary word
  \item \(v\) is the autocorrelation of a word over an alphabet of size \(\geq 2\)
  \item \(v_0 = 1\) and \(v\) satisfies the \textbf{Forward} and \textbf{Backward Propagation Rules}
  \item \(v\) satisfies the predicate \(\Xi\).
  \end{enumerate}
\end{theorem}

Let \(v\in\zon\). We state the original definitions of FPR and BPR.
\begin{definition}
\(v\) satisfies the FPR iff for all pairs \((p,q)\) satisfying \(0\leq
  p<q<n\) and \(v_p=v_q=1\), it follows that \(v_{p+i(q-p)}=1\) for all
\(i=2,\dots,\lfloor (n-p)/(q-p) \rfloor\).
\end{definition}

\begin{definition}
\(v\) satisfies the BPR iff for all pairs \((p,q)\) satisfying \(0\leq
  p<q<2p\), \(v_p=v_q=1\), and \(v_{2p-q}=0\), it follows that
\(v_{p-i(q-p)}=0\) for all \(i=2,\dots,\min(\lfloor p/(q-p)
  \rfloor, \lfloor (n-p)/(q-p) \rfloor)\).
\end{definition}

\end{document}